\documentclass[journal]{IEEEtran}
\usepackage{amsmath,amsfonts,amssymb,amsthm}
\usepackage{epsfig,url}
\usepackage{graphicx,color,footmisc}
\usepackage{subcaption}
\usepackage{cite}
\usepackage{multicol}
\usepackage{multirow}
\usepackage{enumitem}
\usepackage[compact]{titlesec}
\usepackage[ruled,vlined]{algorithm2e}

\SetCommentSty{mycommfont}
\SetKwInput{KwInput}{Input} 
\SetKwInput{KwOutput}{Output} 

\title{5G-Flow: Flexible and Efficient 5G RAN Architecture Using OpenFlow}
\author{\IEEEauthorblockN{Meghna Khaturia\IEEEauthorrefmark{1}, Pranav Jha\IEEEauthorrefmark{1} and Abhay Karandikar\IEEEauthorrefmark{1}\IEEEauthorrefmark{2}} \\
	\IEEEauthorblockA{\IEEEauthorrefmark{1}Department of Electrical Engineering, Indian Institute of Technology Bombay, Mumbai, India\\ Email: \{meghnak, pranavjha, karandi\}@ee.iitb.ac.in} \\
	\IEEEauthorblockA{\IEEEauthorrefmark{2}Director, Indian Institute of Technology, Kanpur, India \\ Email: karandi@iitk.ac.in \\
	\vspace{-27pt}}
	}
   
\begin{document}

\maketitle
\begin{abstract}
   Convergence of multiple access technologies is one of the key enablers in providing diverse set of services to the Fifth Generation (5G) users. Though 3rd Generation Partnership Project (3GPP) 5G standard defines a common core supporting multiple Radio Access Technologies (RATs), Radio Access Network (RAN) level decisions are taken separately across individual RATs as the existing 5G architecture lacks a unified control and management framework for a multi-RAT network. A unified access network is likely to utilize RAN resources more efficiently and provide an improved performance. To bridge these gaps, we present an OpenFlow based RAN architecture comprising of multiple RATs. We refer to it as \textit{5G-Flow}. With minimal changes in the 3GPP 5G RAN and none in the core network, we are able to realize a unified and integrated multi-access 5G-Flow RAN. We simplify the existing 3GPP 5G RAN by replacing RAN nodes with OpenFlow switches and a Software Defined Networking (SDN) controller. Moreover, a UE in the 5G-Flow network can use 5G RAN to connect to any core network (4G or 5G) or directly connect to Internet without going via core network. We also present a simple method to realize 5G non-standalone architecture using 5G-Flow RAN. We have developed an evaluation platform to compare the performance of our architecture with the standard 3GPP 5G network. Results demonstrate significant gains in the network performance of 5G-Flow RAN architecture over existing 3GPP 5G network.   
\end{abstract}

\begin{IEEEkeywords}
Software-Defined Networking, 5G Multi-RAT Network, OpenFlow Protocol and Wireless Network Architecture.
\vspace{-10pt}
\end{IEEEkeywords}
 
\section{Introduction}
   
    The Fifth Generation (5G) cellular standard envisions to support various use-cases, i.e., enhanced mobile broadband, ultra-reliable low latency communications, and massive machine-type communications~\cite{23.501}. These new requirements have led to several innovations and a redesign of the cellular network as part of 3rd Generation Partnership Project (3GPP) 5G standardization, such as service-based architecture, virtualized network functions, control and user planes separation, and network slicing. One of the major advancements is the unification of multiple access technologies in 5G Core Network (5GC), which is an essential step towards enhancing network efficiency. The 3GPP 5G access network incorporates heterogeneous access technologies, i.e., 3GPP access (e.g., next-generation NodeB (gNB), and evolved NodeB (eNB)) and non-3GPP access (e.g., Wi-Fi and wireline access). Towards this, 5G standard has introduced a common interface between 5GC and 5G Radio Access Network (RAN), irrespective of access type, to integrate multiple access networks at 5GC. However, there are certain gaps in the current 3GPP 5G architecture that we discuss next.
   
   Even though 5GC exposes a common interface towards RAN irrespective of the Radio Access Technology (RAT) being used, i.e., N2 interface for control signaling and N3 interface for user data transfer, 5G standard defines different inter-working entities (between RAN and 5GC) for different RATs~\cite{23.501}. For instance, a network entity called Non-3GPP Inter-Working Function (N3IWF) has been introduced for untrusted non-3GPP access (e.g., Wi-Fi), which incorporates N2/N3 protocol stack. In contrast, for 5G New Radio (NR) based access, N2/N3 protocol stack is incorporated within gNB, a new node providing NR data and control planes protocol terminations towards the UE. RAT-specific inter-working entities and associated interfaces result in a complex RAN architecture. Complexity also increases at 5GC, as registration state per RAT is maintained at 5GC if a User Equipment (UE) is connected to more than one RAT. To manage data traffic for a UE using multiple access concurrently, Access Traffic Steering, Switching \& Splitting (ATSSS), an optional feature, has been introduced in 5G standard~\cite{23.501,24.193}. An operator can define policies to flexibly use multiple access for different data flows. However, as 5GC handles the ATSSS feature, it may not optimally perform traffic steering and load balancing across multiple RATs since it is unaware of RAN-level information such as traffic load and radio channel conditions. Another limitation of the existing 3GPP 5G architecture is the close inter-working between RAN and Core Network (CN). While 5GC supports multiple access technologies including Fourth Generation (4G) Long Term Evolution (LTE), it is not possible to use 4G CN with 5G NR gNB. To use 4G CN with gNB, a dual connectivity based approach becomes necessary where a UE is simultaneously connected to eNB and gNB, known as the \textit{non-standalone architecture}. We argue that Software Defined Networking (SDN) based integrated multi-RAT architecture as part of the 5G access network may be better equipped to address these challenges. SDN not only enables the separation of control and data planes through a standardized protocol interface but also supports logical centralization of the network control function through an SDN controller.

   In this paper, we re-architect the 3GPP 5G RAN to realize a unified, integrated, software-defined multi-RAT RAN using OpenFlow Protocol\cite{openflowspec}. We refer to the proposed RAN architecture as \textit{5G-Flow}. It employs OpenFlow switches (network switches based on OpenFlow protocol) and a light-weight OpenFlow controller (also called 5G-Flow controller). Unlike 3GPP 5G RAN design, we propose a unified inter-working entity to interface with 5GC. To realize this entity, we split the protocol stack at RAN nodes (i.e., gNB, N3IWF, etc.), separating the radio protocol stack from the N2/N3 protocol stack. We bring together the radio protocol stack of different RATs and N2/N3 protocol stack as different interfaces of an OpenFlow switch in the multi-RAT network. Similarly, OpenFlow switch at UE decouples Non-Access Stratum (NAS) and IP layers from the underlying radio interface stack at the UE. In essence, the proposed protocol split at UE and RAN nodes along with the usage of OpenFlow switches (to connect the protocol stacks of these entities) decouples UE’s communication with CN from its communication with RAN. This feature allows a UE to flexibly use any radio interface to connect to any CN. For instance, a UE can use 5G NR based radio interface to connect to 4G CN or directly connect to Internet bypassing the CN, which is not possible in the existing 3GPP 5G network.  
    We summarize some of the novel contributions of the work presented in this paper ---
   \begin{itemize}[leftmargin = 0.4cm]      
        \item  One of the key innovations is to vertically split the protocol stack at RAN nodes (at gNB, N3IWF etc.), separating the radio interface from the N2/N3 protocol stack. To enable communication between them, we utilize these protocol stacks as interfaces of an OpenFlow switch. This feature enables us to realize a unified inter-working entity in RAN that replaces RAT-specific inter-working entities.
        \item A light-weight 5G-Flow controller enables RAN level management of downlink as well as uplink dataflows, in turn, utilizing the multi-RAT resources efficiently. Existing OpenFlow abstractions, such as, logical ports and flows are used to control the user traffic flowing through these interfaces. 
        \item UE's connectivity to RAN is fully decoupled from its connectivity to CN. This brings immense flexibility and enables a UE to interface with 5GC, 4G CN, Internet, or any other data network via any 4G/5G/Wi-Fi based RAN. 
        \item 5G-Flow controller manages multiple RATs through OpenFlow protocol messages. The interfaces on the OpenFlow switch are responsible for translating these messages to interface-specific action messages.  
       \item Decoupling of radio and CN protocol stacks at RAN allows us to support the 5G non-standalone architecture in a simplified manner wherein a UE need not be dual connected to gNB and eNB. It can connect to 4G CN via gNB.
       \item UE need not register separately with 5GC even when using multiple RATs concurrently, thereby reducing the complexity of 5GC.
       \item The proposal requires minimal changes in RAN. Moreover, these changes are software-based and can be easily implemented. No changes are proposed in the protocol used between different network entities, i.e., UE-gNB, UE-5GC, or gNB-5GC interface.       
   \end{itemize}

The rest of the paper is organized as follows. In Section II and III, we discuss the literature and background related to the proposed work. In Section IV and V, we discuss the proposed network architecture and its working. We discuss the performance analysis and applications of our proposal in Section VI and VII, respectively. We conclude our work in Section VIII.  

\section{Related Work}
    Several efforts in the literature deal with the application of SDN in cellular networks to make the network more programmable and in turn, more manageable. Research works such as SoftRAN~\cite{softran}, SoftMobile \cite{softmobile}, FlexRAN~\cite{flexran}, SoftNet~\cite{softnet} and Software-Defined Wireless Network (SDWN)~\cite{sdwn} aim at making RAN more programmable, flexible and specifically, use resources efficiently. SoftAir \cite{softair} uses SDN paradigm and hardware abstraction to enable a scalable network architecture for 5G. SoftCell \cite{softcell} aims to introduce scalability and flexibility in cellular core networks. To support rapid protocol evolution, OpenRadio uses SDN paradigm to decouple the wireless protocol into processing and decision planes and provide a programmable interface between the two planes \cite{openradio}. 
    However, none of the above works consider multi-RAT integration in the access network.
    
    We now discuss some research works that propose SDN based network architecture to manage heterogeneous RATs. In \cite{orchestra}, the authors introduce a virtual MAC layer at network nodes and users in order to manage heterogeneous RATs in a technology independent manner. 5G-EmPower \cite{empower} proposes an SDN based framework to manage heterogeneous access networks by abstracting the technology-dependent features of RATs. In \cite{openroads,openroads2}, the authors have developed a prototype that augments WiMAX and Wi-Fi APs with OpenFlow. OpenRAN \cite{openran} uses virtualization to design a software-defined multi-access RAN. However, extension of the proposed concept in \cite{orchestra,empower,openroads,openroads2,openran} for 3GPP 4G/5G network is not straightforward. 
    
    The authors in \cite{sdn_akshatha} propose an end-to-end SDN based architecture for a mobile network with a controller managing both access and CN data plane elements by providing an abstract view of the underlying network resources to the controller. In~\cite{open5g}, the authors propose a multi-RAT architecture and define a unified and open interface between the controller and multi-RAT data plane.    In \cite{multiratmc}, the authors propose a convergence sub-layer over layer 2 of multiple RATs in order to tightly integrate them. While \cite{sdn_akshatha,open5g,multiratmc} modify the radio protocol stack for RAT integration, the radio stack of different RATs in 5G-Flow RAN remains unchanged. To integrate LTE with Wi-Fi, LTE-Wireless Local Area Network Aggregation (LWA) mechanism has been proposed by 3GPP \cite{lwa1,lwa2}. LWA proposes an emulation layer over Wi-Fi AP that encapsulates LTE packets in Wi-Fi MAC frame. 
    
    None of the above works propose enhancements to the recently defined 3GPP 5G RAN architecture to the best of our knowledge. Moreover, use of OpenFlow to propose a unified inter-working entity and thereby integrating multiple RATs (with no changes in radio stack) in downlink and uplink has also not been discussed in literature. Decoupling UE's communication with RAN from its communication with CN within 3GPP 5G framework is also one of the novel contributions of our work. 
    

\section{Background}
In this section, we discuss OpenFlow protocol and some basic 3GPP 5G terminologies which form the basis of our discussion ahead.

    \textbf{OpenFlow Protocol}:  OpenFlow (OF) is a protocol used by an SDN Controller to manage the forwarding plane of a network.~\cite{openflowspec}. An OF compliant network consists of a logically centralized OF controller (SDN controller) and multiple OF switches (forwarding plane). An OF switch supports an OF client, which communicates with an OF controller through a secure channel. The controller manages the flow tables present in OF switches using OpenFlow protocol. A flow-table comprises of several flow-entries that match on a packet based on match-fields such as IP address or TCP port number. Based on the flow-entry, an action is taken (e.g. forward or drop) on the matched packet. OF switch supports physical and logical ports. A physical port corresponds to a hardware interface on an OF switch. A logical port does not directly correspond to a hardware interface, but it is an abstraction of a physical port. It can be used to implement complex processing of packets in OF switch. OF-Configuration (OF-Config) protocol, a complementary protocol based on Network Configuration Protocol (NETCONF) \cite{netconf}, helps in configuration and management of OF switches~\cite{openflowconfig}. OF-Config is responsible for association between a controller and an OF switch and configuration of physical and logical ports. To inform the OF controller about various events at OF switches such as link failures or configuration changes, OpenFlow Notifications Framework provides a method to subscribe to asynchronous messages based on predefined events~\cite{openflownotifications}.

\begin{figure}
        \centering
        \includegraphics[width=0.7\linewidth,trim={0cm 10cm 13cm 0cm},clip]{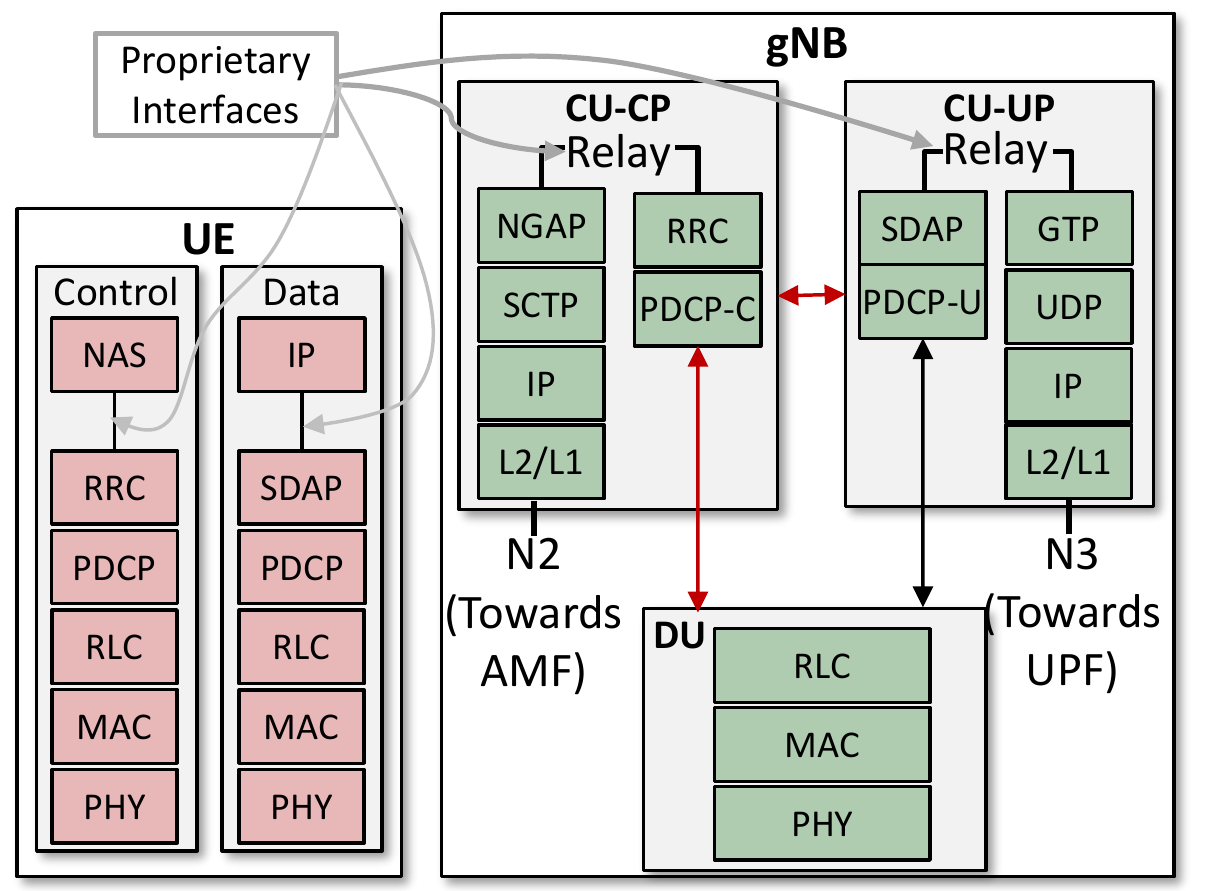}
        \caption{3GPP 5G gNB and UE Protocol Stack}
        \label{fig:ue_gnb_stack}
    \end{figure}

    \textbf{5G Primer}: Fig.~\ref{fig:ue_gnb_stack} illustrates the protocol stack of UE and gNB as defined in the 3GPP 5G standard~\cite{23.501}. gNB consists of a Centralized Unit (CU) and one or many Distributed Units (DUs). gNB-CU can be further divided into control plane (gNB-CU-CP) and user plane (gNB-CU-UP) components. UE and gNB both have Radio Resource Control (RRC) layer that facilitates control plane interaction between the two entities. RRC also helps in radio connection establishment and release, mobility management, and setting up of radio bearers. Service Data Adaptation Protocol (SDAP) layer, along with the underlying protocol stack, at UE and gNB, is responsible for user plane data transfer over the radio interface along with Quality of Service (QoS) handling of dataflows. NAS Layer present at UE is responsible for non-radio related signaling between UE and 5GC. 5G RAN communicates with 5GC through N2 and N3 interfaces. Next-Generation Application Protocol (NGAP) layer, together with the underlying protocol stack (N2 interface), is responsible for all signaling (control) message exchange between RAN and Access and Mobility Management Function (AMF) in 5GC. Data packets between RAN and User Plane Function (UPF)  are exchanged over N3 interface using GPRS Tunneling Protocol (GTP) and the underlying UDP/IP protocol stack. 
    
    The NGAP-RRC and SDAP-GTP protocol interfaces are tightly coupled with each other and utilize proprietary vendor specific communication mechanism, as shown in Fig.~\ref{fig:ue_gnb_stack}. Though not shown in the figure, other 5G RAN nodes such as N3IWF also use proprietary interfaces for communication between their radio and CN stacks. This tight coupling between radio and CN protocol stacks on RAN nodes is one of the reasons behind RAT specific Inter-working functions in 5G. Similarly, higher layers (NAS and IP Layers) at UE are tightly coupled to the underlying radio protocol stack, leading to close inter-working between RAN and CN.
    
    When a UE wants to connect with a data network, such as Internet, the 5G network establishes an end-to-end tunnel between UE and UPF (i.e., a Protocol Data Unit (PDU) Session).  Additionally, a unique signaling link between UE and 5G Network (both at 5GC and RAN) is also established to exchange control messages between the UE and the 5G network. The end-to-end data tunnel takes the form of a Data Radio Bearer (DRB) at the air interface, while the signaling link takes the form of a Signaling Radio Bearer (SRB). Radio bearers are essentially layer 2 tunnels. Between gNB and 5GC (AMF+SMF), UE specific signaling messages (NAS messages) are exchanged through a (per UE) unique association over N2 interface, whereas a PDU Session takes the form of a GTP tunnel between gNB and 5GC (UPF).

    In 3GPP 5G network, different unique identifiers are used to identify UE specific signaling association and data over different interfaces. For example, to uniquely identify UE over N2 interface, gNB uses RAN UE NGAP ID, whereas AMF uses AMF UE NGAP ID. Similarly, UE specific data sessions are uniquely identified via GTP Tunnel End-point Indicator (TEID) on N3 interface.

 \begin{figure*}
    \begin{subfigure}{.49\textwidth}
      \centering
      \includegraphics[width=0.95\linewidth,trim={0 8.5cm 11cm -0.25cm},clip]{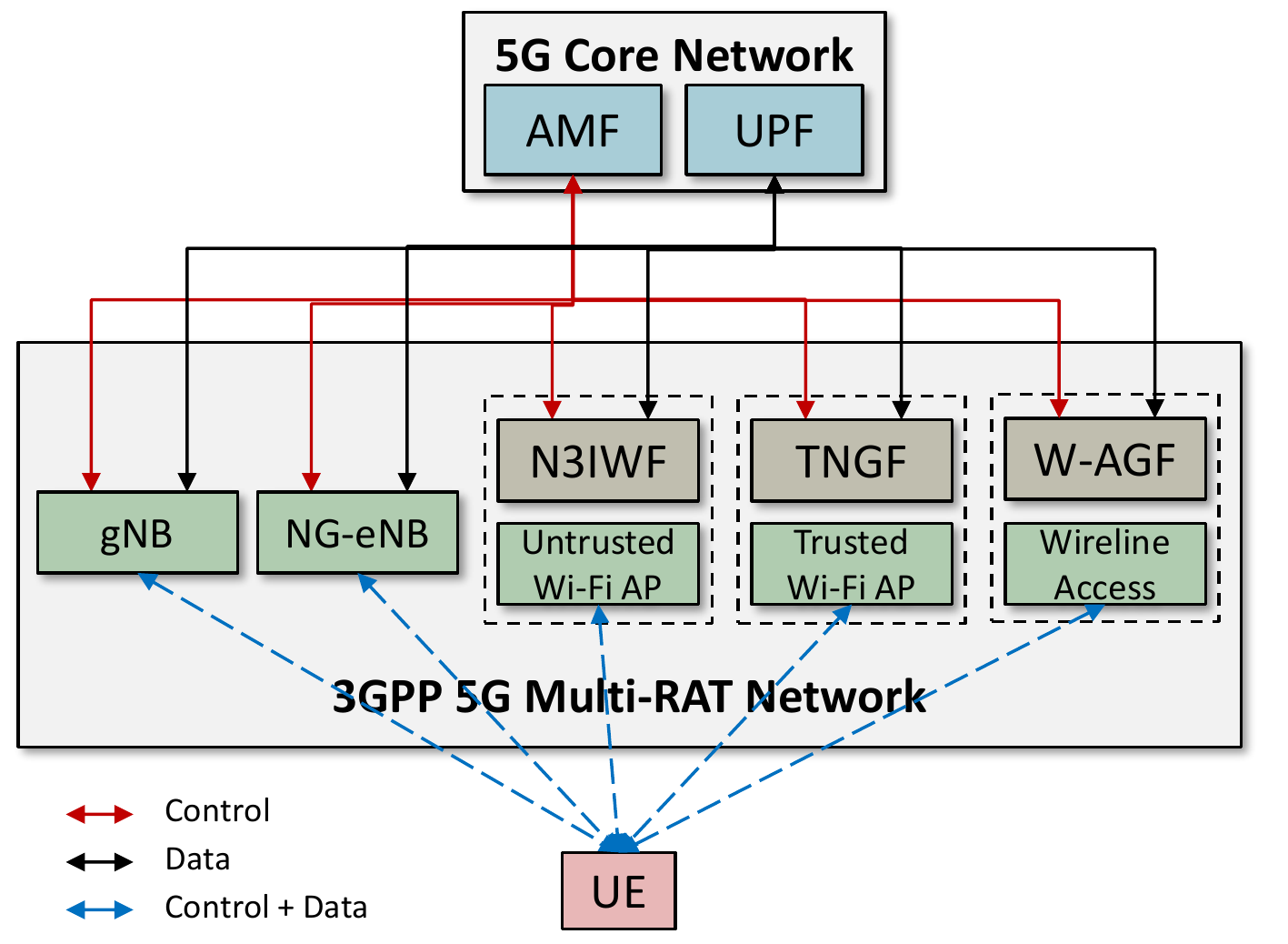}
      \caption{3GPP 5G multi-access RAN architecture}
      \label{fig:5g}
    \end{subfigure}
    \begin{subfigure}{.49\textwidth}
      \centering
      \includegraphics[width=0.95\linewidth,trim={0cm 8.25cm 11cm 0cm},clip]{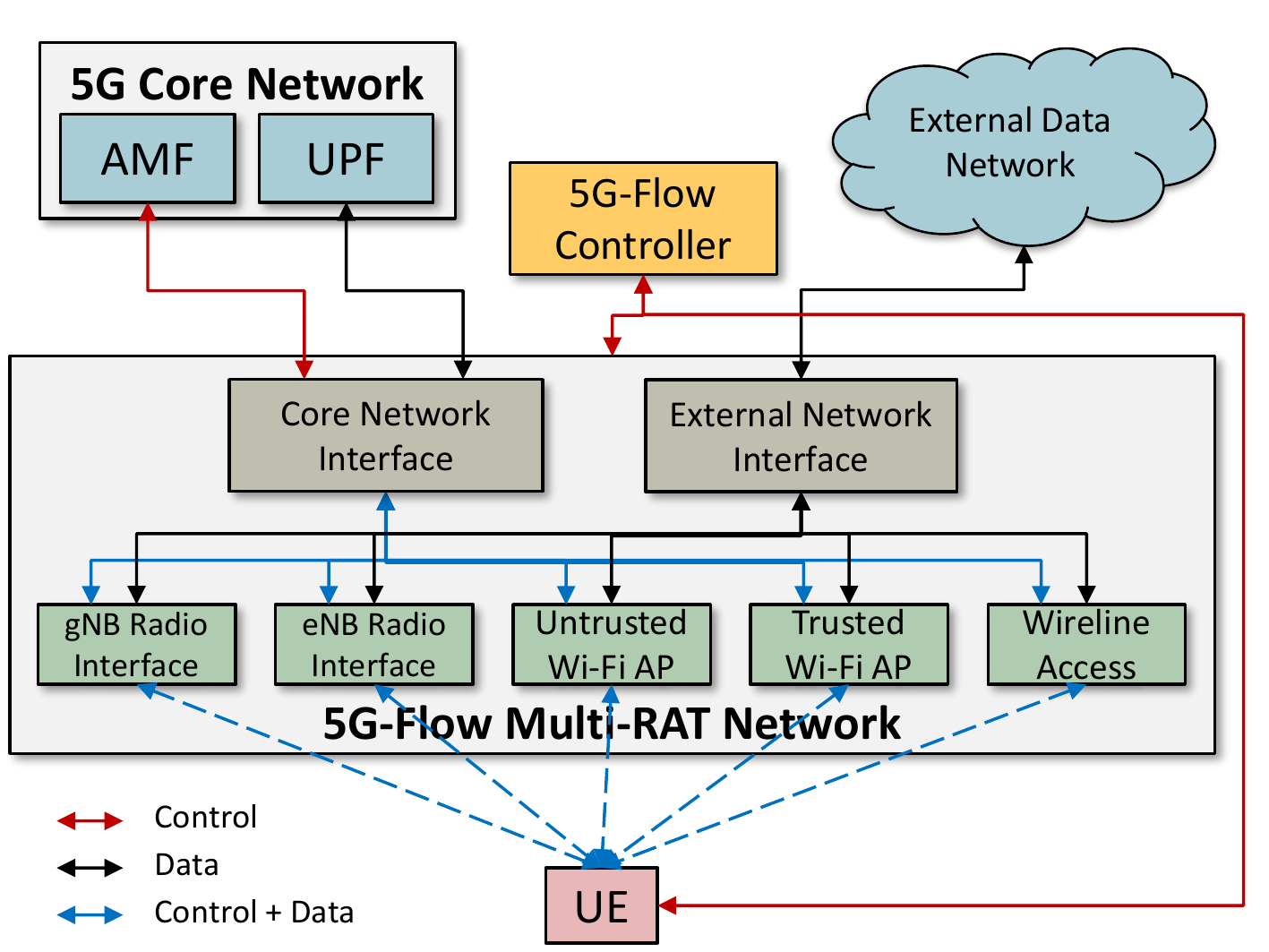}
      \caption{Proposed 5G-Flow RAN Architecture}
      \label{fig:5gflow}
    \end{subfigure}
    \caption{Conceptual diagram for 5G-Flow RAN architecture and its comparison with the current 3GPP 5G RAN}
    \vspace{-12pt}
\end{figure*}
\section{5G-Flow RAN Architecture}
    The current 3GPP 5G multi-RAT RAN architecture is illustrated in Fig.~\ref{fig:5g}. The figure shows how various access technologies interface with 5GC using separate inter-working entities. Untrusted Wi-Fi access uses N3IWF, trusted Wi-Fi access uses Trusted Non3GPP Gateway (TNGF), and wireline access uses Wireline Access Gateway Function (W-AGF) to interface with 5GC. Fig.~\ref{fig:5gflow} illustrates our proposed multi-RAT RAN architecture.  A multi-RAT 5G-Flow RAN communicates with 5GC through a \textit{unified inter-working entity} instead of separate inter-working functions. Additionally, we aim to enable a flexible interface between RAN and CN, such that any RAN can connect with any cellular CN or Internet directly. The software-defined \textit{5G-Flow controller} acts as a multi-RAT RAN controller that manages the unified inter-working entity and dataflows across multiple RATs in RAN. Since the controller has access to RAN-level information such as traffic load and radio channel conditions, it can efficiently manage the downlink dataflows across RATs. As shown in Fig.~\ref{fig:5gflow}, the controller also controls the UE, which enables uplink dataflow management in a multi-RAT RAN. To realize the proposed 5G-Flow RAN architecture, we apply the OpenFlow concepts~\cite{openflowspec}. We envision 5G RAN as an OF network comprising of a 5G-Flow controller (as OF controller) and OF switches that are instantiated on the network side and the associated UEs, as shown in Fig.~\ref{fig:ofswitchblock}. We discuss the modifications to the current 3GPP 5G architecture to realize the 5G-Flow network in the remainder of this section.

    \subsection{Proposed Modifications at the multi-RAT Network}
\begin{figure*}
        \begin{subfigure}[t]{0.35\textwidth}
            \centering
            \fbox{\includegraphics[width=0.96\textwidth,trim = {0 3cm 13.2cm -0.9cm},clip]{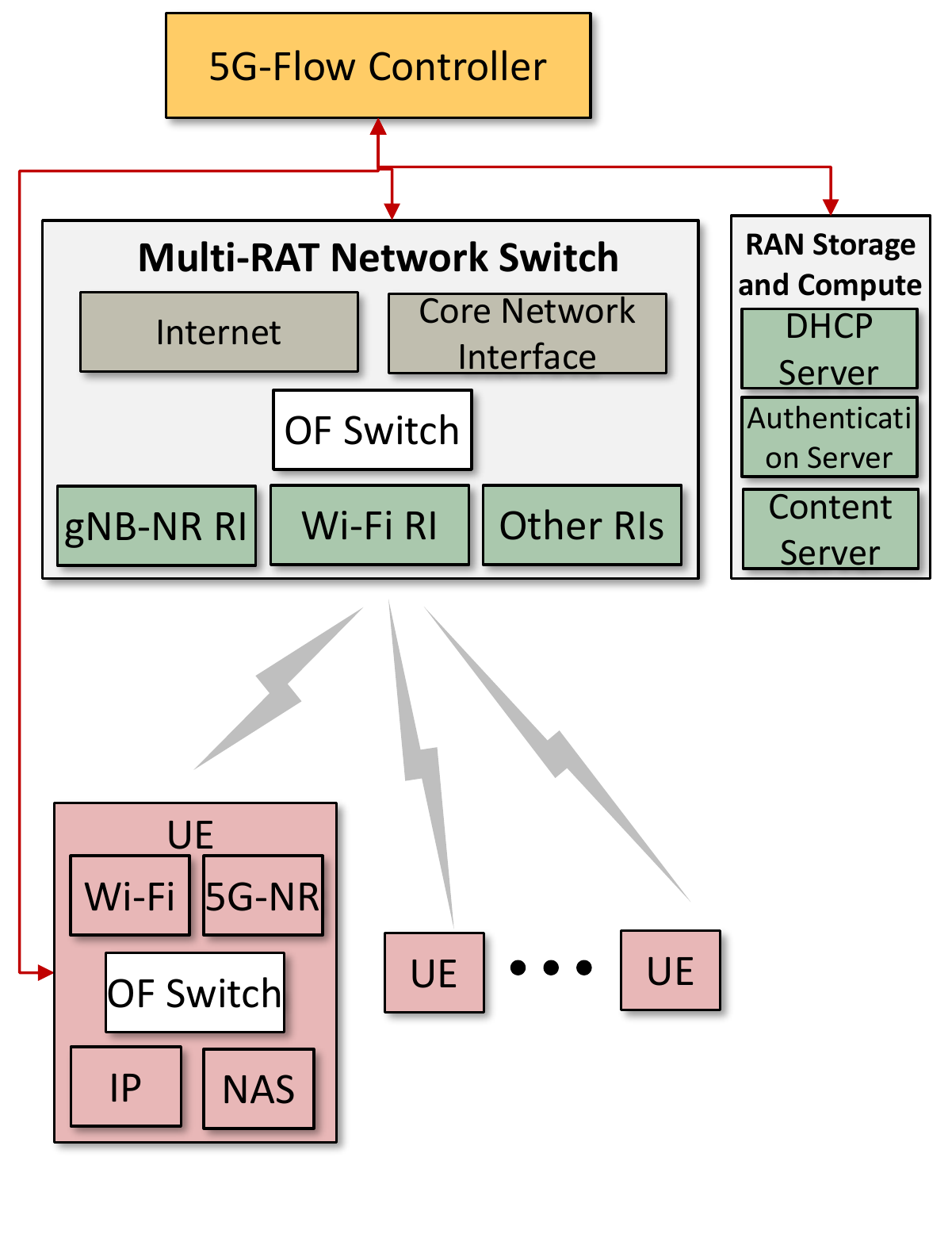}}
            \caption{Block Diagram of 5G-Flow RAN}
            \label{fig:ofswitchblock}
        \end{subfigure}%
        \quad
        \begin{subfigure}[t]{0.64\textwidth}
            \centering
            \fbox{\includegraphics[width=0.95\textwidth, trim = {0cm 3cm 4.5cm 0cm}, clip]{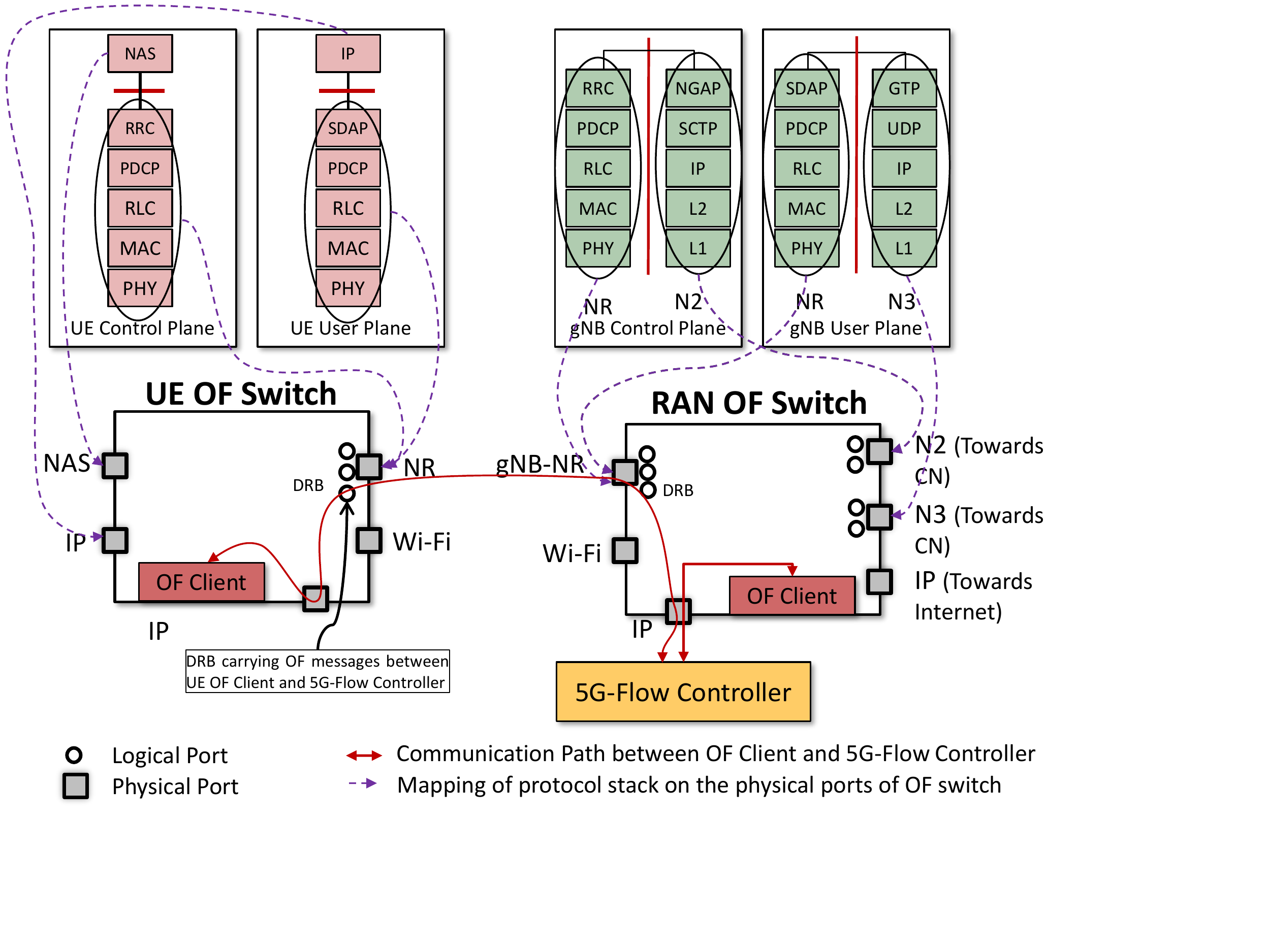}}
            \caption{Application of OpenFlow Protocol at RAN}
            \label{fig:ofswitch}
        \end{subfigure}
        \caption{Implementation of 5G-Flow RAN Architecture using OpenFlow Protocol}
        \vspace{-12pt}
    \end{figure*}

        The existing 3GPP 5G RAN consists of various multi-RAT network nodes, including 3GPP access (e.g., gNB, eNB) and non-3GPP access (e.g., Wi-Fi, N3IWF). To integrate multiple RATs in 5G-Flow RAN and enable a unified inter-working entity, we propose a protocol split between radio interface and N2/N3 protocol stack of RAN nodes. For 3GPP access nodes such as gNB, split happens at the gNB node itself, whereas for non-3GPP access, it is done at the inter-working function such as N3IWF. To illustrate the protocol split, we take an example of gNB. As discussed previously, gNB consists of tightly coupled NR protocol stack (which interfaces with a UE) and N2/N3 protocol stack (which interfaces with the 5GC). As shown in Fig.~\ref{fig:ofswitch}, we split gNB vertically and separate gNB-NR and N2/N3 protocol stack. We can similarly split N3IWF node. We now introduce an OF switch, referred to as Multi-RAT Network (MRN) OF switch, which is responsible for bridging radio and N2/N3 protocol stacks of multiple RATs. These protocol stacks form different interfaces (physical ports) of MRN OF Switch. The 5G-Flow controller directs MRN OF switch to process the messages from different radio interfaces and deliver them to N2/N3 stack towards 5GC and vice-versa. This way, the MRN OF switch along with the 5G-Flow controller replaces all RAT-specific entities such as gNB, eNB, N3IWF, etc. and exposes a unified interface towards CN. 

        The MRN OF switch has physical ports both at the radio and the 5GC interface side, as shown in Fig.~\ref{fig:ofswitch}. Both the control plane (RRC and underlying protocol stack) and the data plane (SDAP and underlying protocol stack) of gNB-NR radio interface map to one of the radio side ports of the OF switch. Similarly, Wi-Fi Media Access Control (MAC) and physical layer map to another port.  NGAP and GTP protocol layers (along with the underlying N2/N3 stack) map to the physical ports on the 5GC side. The physical port, labeled as IP,  interfaces with the external data network. As the radio interface stack is decoupled from the N2/N3 protocol stack, OF switch can steer the data traffic of a UE towards IP port, enabling direct connectivity with Internet bypassing the CN. An interface towards 4G CN is not shown in the figure, but it can be easily incorporated by adding S1-Mobility Management Entity (MME) interface as a separate physical port in the proposed OF switch. This feature enables a UE with 4G compatible NAS layer to communicate with 4G CN via 5G RAN. This feature is discussed in the detail in Section~\ref{sec:internet}.

        In Fig.~\ref{fig:ofswitch}, we have mapped Wi-Fi MAC layer to the radio side port in MRN OF switch. However, the user plane of N3IWF uses additional protocol layers i.e., Generic Routing Encapsulation (GRE) and IP Security (IPsec), for creating a secure tunnel between UE and N3IWF over Wi-Fi radio interface. Similarly, it uses TCP and IPsec protocol layers in the control plane for encapsulating NAS messages. Our architecture provides flexibility in employing these protocols. If a UE does not want to use GRE and the underlying protocol stack for some dataflows, a logical port can be created at UE and RAN, which transparently passes the data packets through Wi-Fi interface without any processing. Further, a logical port enabled with GRE, and IPsec protocol layers can be created over Wi-Fi MAC based physical port if a UE needs a secure tunnel for another data flow. 

    \subsection{Proposed Modifications at UE}
        An OF switch, introduced at UE, decouples NAS (that communicates with 5GC) and IP protocol layers from the underlying radio protocol stack, as shown in Fig.~\ref{fig:ofswitch}. We introduce a common IP layer instead of RAT-specific IP layers. There can be different radio interface stacks depending on the technology, but NAS and IP layers remain common. At the radio side ports of UE OF switch, NR (RRC/SDAP, and the underlying protocol stack) and Wi-Fi (MAC and physical layer) radio stacks are mapped. 

        The UE OF switch, along with the 5G-Flow controller, manages the UE's radio connectivity and enables uplink dataflow management across multiple RATs. Moreover, when a UE is connected to 5GC via multiple RATs, it registers only once. In the existing 3GPP architecture, a UE connected to 5GC through more than one RAT has to register with 5GC separately via each RAT. With the separation of NAS layer from radio protocol stack (in addition to radio and N2/N3 protocol split at the network side), UE's communication with RAN is completely decoupled from its communication with 5GC. This feature allows a UE to flexibly connect to different networks such as 5GC, 4G CN, or directly to Internet. 

    \subsection{OF Switch Configuration}
    \label{sec:wifi}
        As discussed before, OF controller can create and configure logical ports on an OF switch using OF-Config protocol. The meaning (processing) associated with these logical ports varies according to the underlying interface being used. On the gNB-NR interface at the MRN OF switch, a logical port represents a UE-specific radio bearer. At least two logical ports are created for each UE, one for SRB and another for DRB. To identify these logical ports uniquely, we use SRB/DRB ID as defined in 5G standard. A logical port at NGAP interface signifies a UE-specific NGAP association and is identified by RAN UE NGAP ID. Similarly, at GTP interface, a logical port implies a PDU session of a UE and is identified by a GTP-TEID.

        At NR interface of the UE OF switch, a logical port represents SRB and DRB  of a UE (similar to MRN OF switch). At IP interface, logical ports signify ongoing PDU sessions for a UE. 

        Despite the differences in processing at each of the physical interfaces, usage of logical ports provides a uniform abstraction to be used by 5G-Flow controller to configure flow paths through the switch. The 5G-Flow controller simply configures logical ports on the physical interfaces in an OF switch. It is the responsibility of the interface to translate OF-Config message (for port creation) to interface specific handling. For instance, a DRB, along with a GTP tunnel, needs to be established during a PDU Session setup. When the gNB-NR interface receives a message to create a new logical port corresponding to a DRB, the RRC layer on the gNB-NR interface translates it to configure its underlying lower layers, e.g., PDCP, RLC layers for the local DRB configuration. It also exchanges \textit{RRC Reconfiguration} messages with the UE for a corresponding DRB configuration on the UE. Similarly, a logical port creation message sent to N3 interface is translated (by the interface) to create a GTP tunnel. A logical port creation message to N2 interface gets translated into the creation of a unique UE-specific NGAP association with 5GC.

\begin{figure}
    \centering
    \includegraphics[scale=0.5,trim = {0 8.5cm 5cm 0},clip]{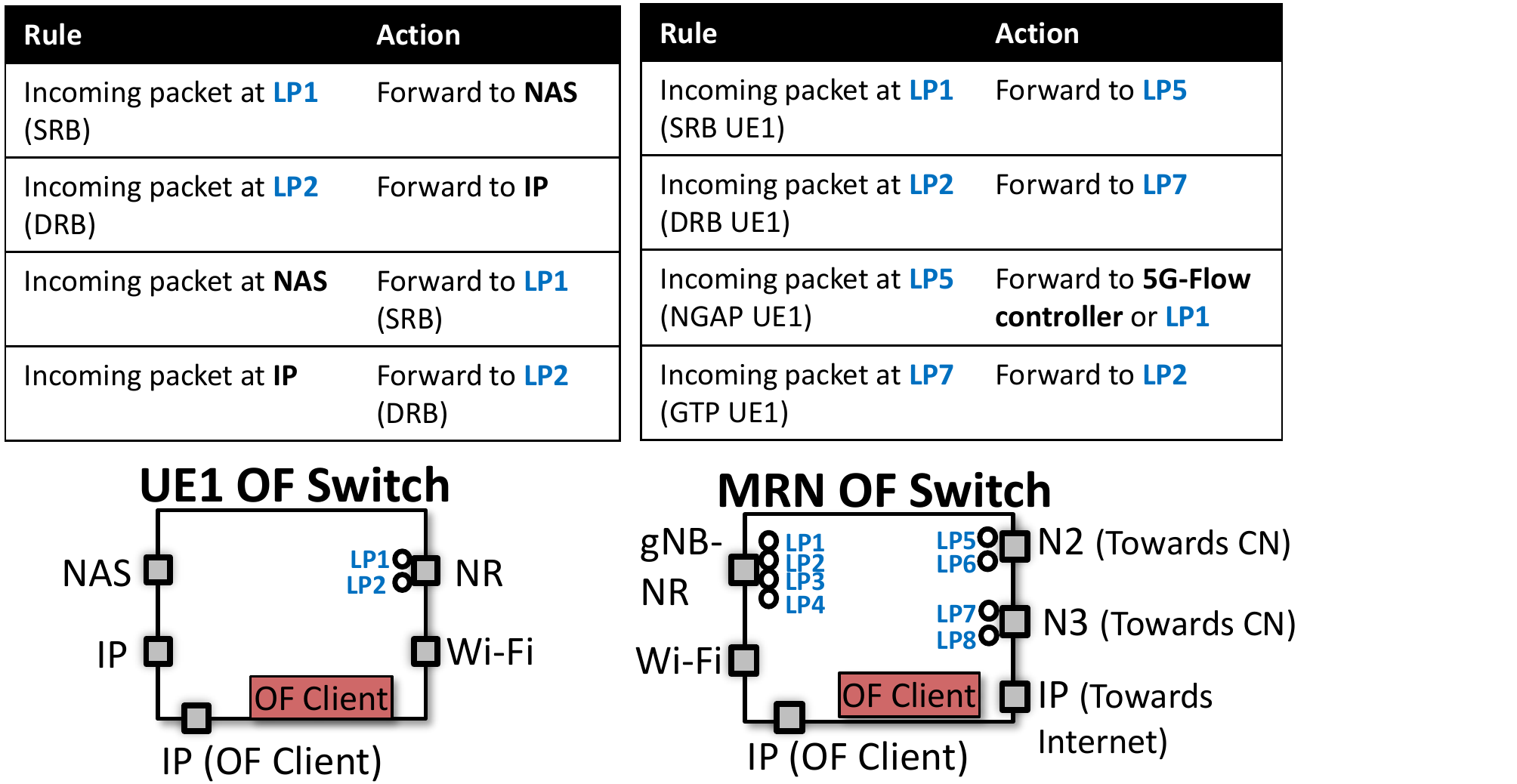}
    \caption{Example of flow entries at UE and MRN OF switch and their relation with Logical Ports (LPs) in the switches}
    \label{fig:flowtable}
\end{figure}

        Once the logical ports are created on the interfaces, 5G-Flow controller defines a mapping between different logical ports across the interfaces on the OF switch, as shown in Fig.~\ref{fig:flowtable}. For instance, a mapping between DRB and GTP tunnel is created for a UE, or a mapping between UE-specific NGAP association and SRB is created. The controller realizes these mappings through flow entries added at the OF switches. These port mappings enable the simple forwarding of data and signaling messages through the OF switch and also makes the control and management task for the 5G-Flow controller easy.

        A key novelty of the proposal is the usage of GTP tunnels, radio bearers, or UE-specific NGAP association as logical ports. These entities (a radio bearer, or a UE-specific NGAP association) carry a specific set of data flows or messages. By using them as logical ports, we can virtualize them and enable their manipulation by an OF Controller through OF protocol.

\section{Working of the 5G-Flow RAN}
We describe various system procedures in this section that are important for understanding the working of our architecture.
\subsection{Initial Connection Setup}
\begin{figure}
    \centering
    \includegraphics[scale=0.58]{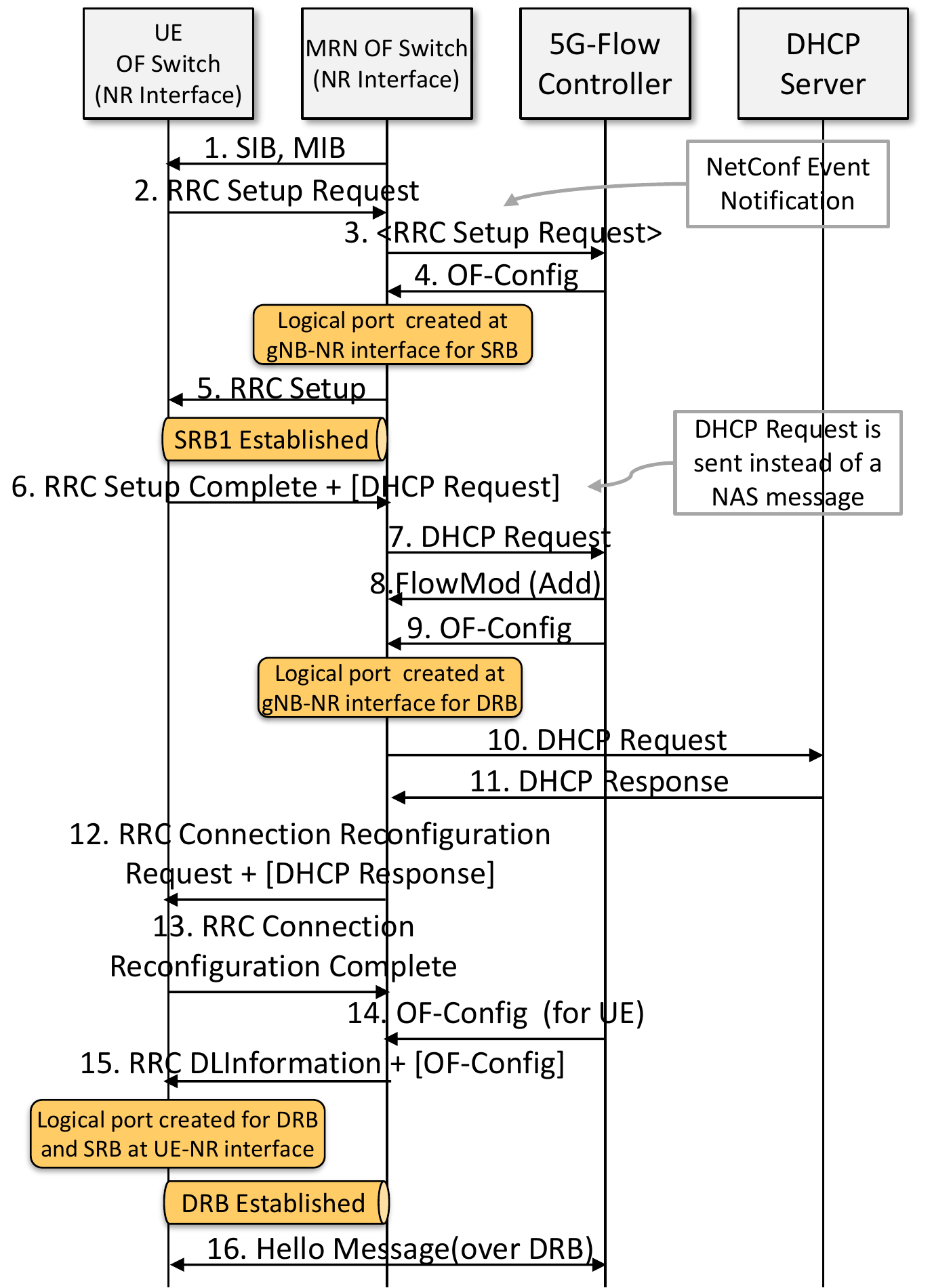}
    \caption{Initial Connection Setup}
    \label{fig:callflow}
\end{figure}

In this section, we discuss how OF switches can communicate with the 5G-Flow controller. The control messages between an OF switch and the controller are exchanged over a TCP/IP connection. The MRN OF switch and the 5G-Flow controller are co-located and can communicate over a wired interface. Fig.~\ref{fig:callflow} shows how communication between an UE OF Client (UE OF switch) and the 5G-Flow controller is established. Here, we have assumed that the default path for UE-controller communication is via gNB-NR interface, but the path can be established via Wi-Fi radio interface as well. We explain the call flow to set up the initial connection next.

\begin{itemize}[leftmargin = 0.4cm]
    \item To establish a radio connection with gNB-NR interface, UE sends an \textit{RRC Setup Request} over common channel (SRB0). Since 5G-Flow controller is responsible for admission control, gNB-NR interface notifies the controller via a NETCONF notification so that a decision for UE admission can be taken. If admitted, the 5G-Flow controller sends an OF-Config message to create a logical port on gNB-NR interface for the subsequent signaling messages exchange (via SRB1) with the UE. gNB-NR interface sets up SRB1, maps it to the logical port, and sends \textit{RRC Setup}. As the UE OF client still does not have a TCP/IP connection with the controller, a logical port can not be created at UE using OF-Config. Instead, UE uses the default physical port mapped to 5G-NR interface for initial signaling.
    \item UE responds with \textit{RRC Setup Complete} message and a \textit{DHCP request} is sent in NAS message field instead of \textit{Registration Request}. gNB-NR interface forwards \textit{DHCP request} towards the 5G-Flow controller as a table-miss is triggered at MRN OF switch. The controller upon receiving this message sends a FlowMod (Add) command to add the flow entry at MRN OF switch. 5G-Flow controller also sets up a logical port at gNB-NR interface for DRB of the UE, using OF-Config message. This DRB is created for carrying OF client messages from UE to the 5G-Flow controller.
    \item gNB-NR interface forwards \textit{DHCP request} to the DHCP server which processes the request. The DHCP response is sent to UE in the NAS message field of \textit{RRC Connection Reconfiguration Request}. A DRB has now been established. 
    \item An OF-Config message is sent to the UE to configure a logical port and map it to the newly established DRB. This message creates a logical port for SRB as well, which is used for future signaling messages. The communication path between the UE OF switch and the 5G-Flow controller is now established via DRB, and they exchange \textit{Hello} message over this DRB.
\end{itemize}

This callflow illustrates how UE's communication with RAN is decoupled from its communication with CN. UE can use RAN to exchange DHCP and OpenFlow messages with entities located in the edge instead of exchanging NAS messages and data with the CN.

\subsection{Registration and PDU Session Setup with 5GC}

\begin{figure}
    \centering
    \includegraphics[scale = 0.6, trim = {0cm 4.4cm 11.3cm 0cm }, clip]{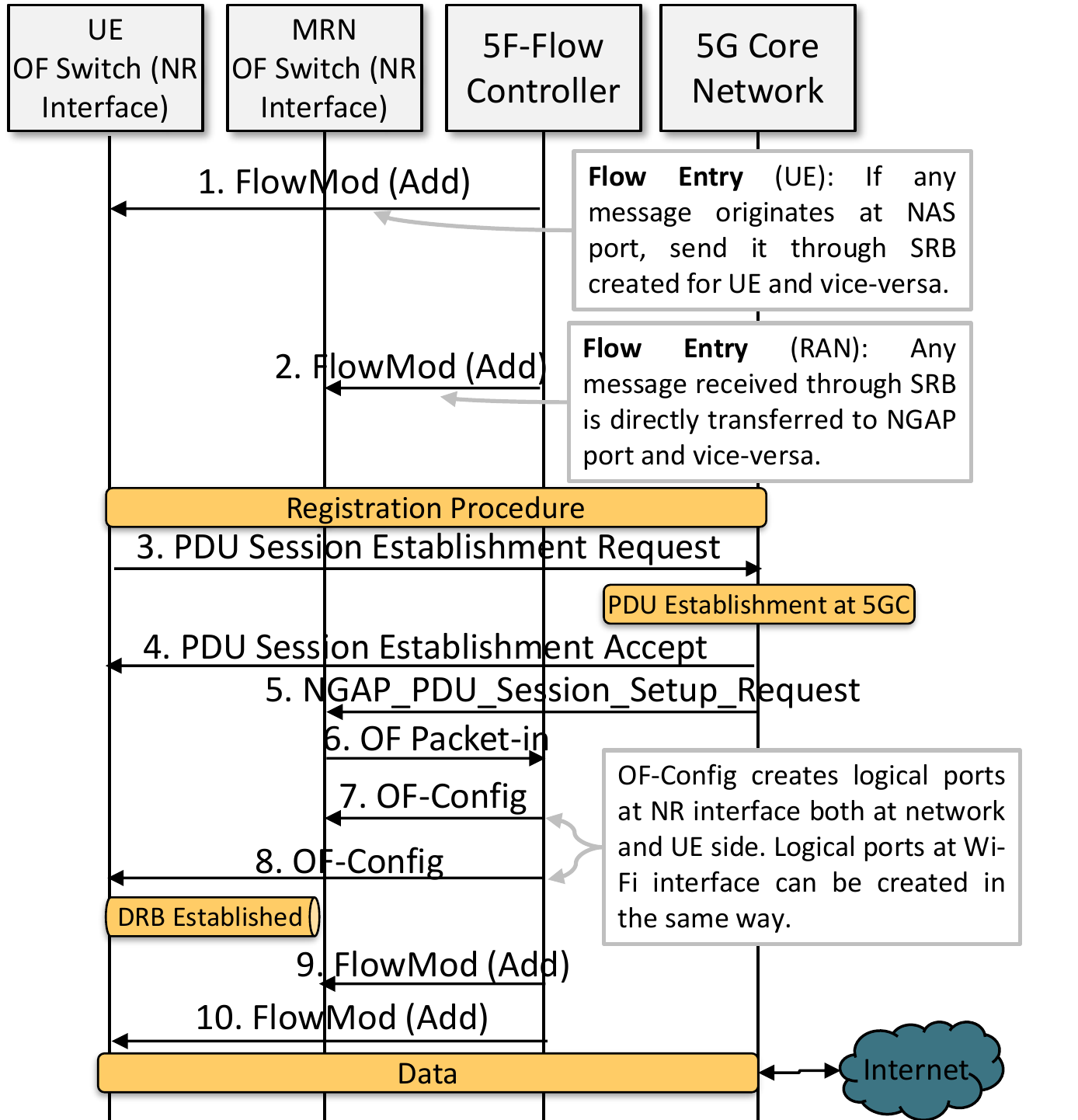}
    \caption{Call Flow to illustrate the communication of UE with 5G Core Network}
    \label{fig:callflow_5G}
\end{figure}
If a UE wants to access the 5G cellular network, it needs to register with the 5GC. Here, we take an example of how a UE registers via gNB. The registration of a UE can also take place via a Wi-Fi network, if available. To facilitate UE's communication with the 5GC, 5G-Flow controller can proactively set up a path that delivers NAS messages from a UE to the 5GC and vice-versa, as shown in steps 1 and 2 in Fig.~\ref{fig:callflow_5G}. It can also be implemented through a reactive method in which the 5G-Flow controller adds the flow entries after the first NAS message originates at UE, and a table-miss is detected. 

Next, we discuss how a PDU session is established via 5G-Flow RAN. When AMF (in 5GC) receives the NAS message (\textit{PDU Session Establishment Request}) from a UE, it informs SMF, which creates a PDU session for the UE. SMF sends PDU session related information to AMF, i.e., GTP tunnel end-point and QoS information. AMF forwards this information as an NGAP message to RAN along with a NAS message (\textit{PDU session Establishment Accept}). The MRN OF switch transparently forwards the NAS message to the UE. However, unlike NAS message, it processes the NGAP message and forwards PDU session related attributes to the 5G-Flow controller via \textit{OF Packet-in} message. The controller, based on the Packet-in message and RAT-specific information, decides how the incoming dataflow should be distributed among the available RATs. It, then, sends OF-Config messages to the radio interfaces to create logical ports, which are interpreted by these interfaces for configuration of the underlying protocol stack. For instance, configuration of lower layers by RRC to create a DRB when a logical port creation message is received at gNB-NR interface. Also, an OF-config message is sent to GTP interface of the MRN OF switch for creating a logical port which signifies GTP tunnel for a UE's PDU Session.  After this, flow entries are added at the UE OF switch and the MRN OF switch that maps the newly created DRB to IP port at UE and maps the DRB to the GTP tunnel, respectively. The path for UE PDU session is now set up.


\subsection{Dynamic Dataflow Management}
\label{sec:flow}

\begin{figure}
    \begin{subfigure}{0.5\textwidth}
      \centering
      \includegraphics[scale=0.6,trim={0 11.8cm 13cm 0.22cm},clip]{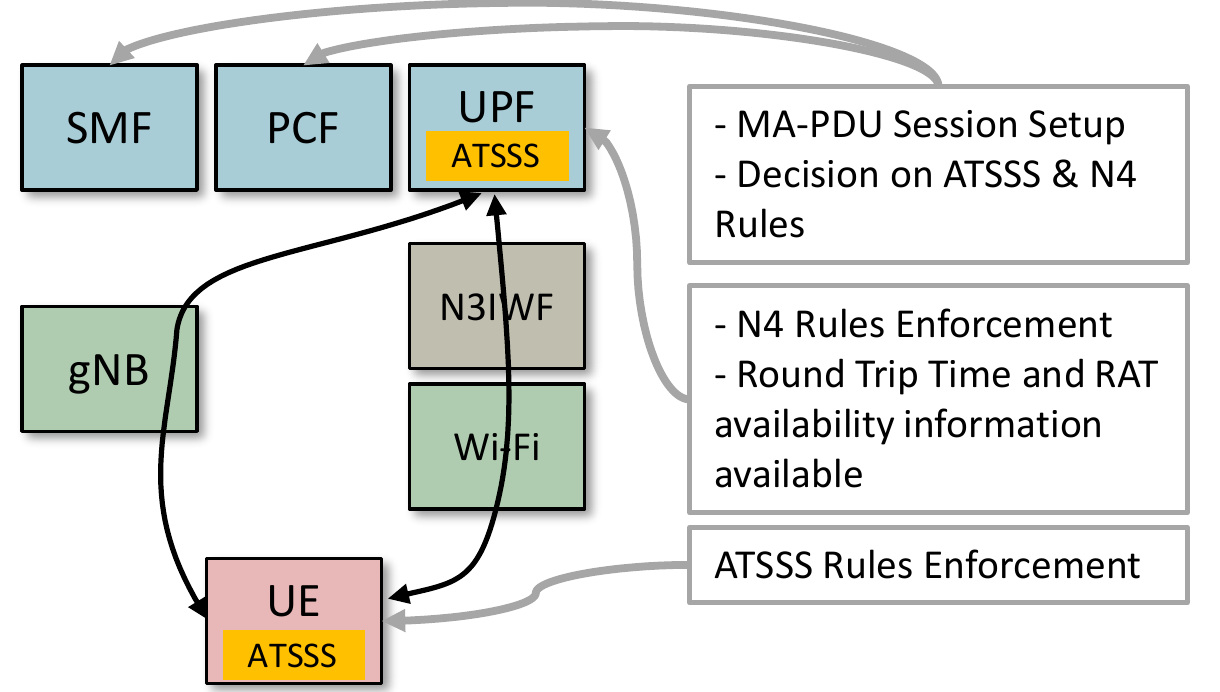}
      \caption{Dataflow in 5G Network under ATSSS feature}
      \label{fig:5gdataflow}
    \end{subfigure}%
    \quad
    \begin{subfigure}{0.5\textwidth}
      \centering
      \includegraphics[scale=0.6,trim={0 13.2cm 14cm 0cm},clip]{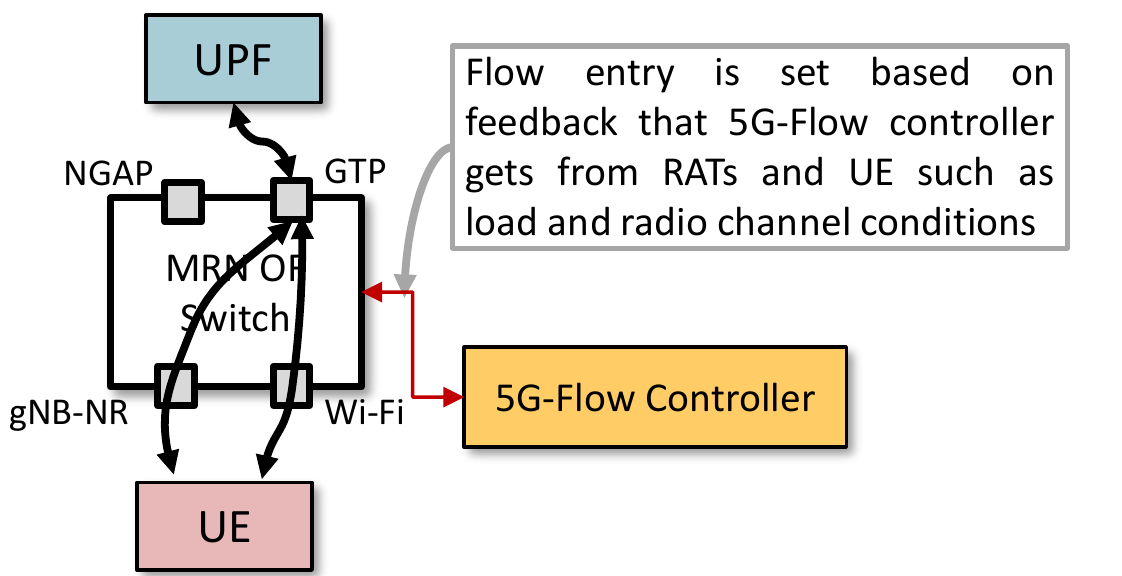}
      \caption{Dataflow in 5G-Flow Network}
      \label{fig:dataflow}
    \end{subfigure}
    \caption{Comparison of dataflow with 5G ATSSS feature and dataflow with 5G-Flow architecture}
\end{figure}

5G has introduced ATSSS feature, which manages the multi-access traffic in the 5G network. Fig.~\ref{fig:5gdataflow} illustrates how the multi-access traffic flows through the 3GPP 5G network~\cite{23.501,24.193}.  UE can initiate a Multi-Access-Protocol Data Unit (MA-PDU) session to enable PDU exchange between UE and UPF via 3GPP and non-3GPP RATs simultaneously. To manage the uplink traffic, UE considers the ATSSS rules provided by the 5GC. To manage the downlink traffic, UPF considers ATSSS rules provided by SMF over N4 interface along with the feedback information from UE. However, the feedback information available at UPF is limited to only Round Trip Time (RTT) and RAT availability. This information may not be sufficient to route the flow through multiple RATs optimally.

As shown in Fig.~\ref{fig:dataflow}, the flow configuration in the proposed 5G-Flow network happens at the RAN-level, where utilizing various RAT-specific attributes is viable. The controller can periodically access the value of specific attributes, such as traffic load at different RAT interfaces, flow statistics, and radio signal strength, to optimize the dataflow. The first two attributes are available at the OF switch. To access the radio signal measurement available at RRC (gNB-NR), the controller can subscribe to the measurement data at the OF switch via OF-Config (NETCONF) notifications~\cite{openflownotifications}. The asynchronous notification mechanism is supported by OF-Config protocol and it allows us to define notification messages, in addition to the already available set of notifications. 

Based on these parameters, an optimized policy for flow management can be determined, and the incoming flow is split across available RATs, as shown in the figure. We also have the flexibility to decouple uplink and downlink in 5G-Flow RAN. As the controller manages MRN and UE OF switch separately and can add distinct flow entries in both the switches, uplink and downlink for a user can be easily decoupled. We evaluate the performance of data flow management in the 5G-Flow network viz-a-viz the 3GPP 5G network in the next section. 

    \section{Performance Analysis}
    \label{sec:performance}
    To analyze the performance of our architecture, we have built an evaluation platform using MATLAB\cite{matlab}. An open-source 5G simulator with a fully developed protocol stack is not yet available, so we have developed a 5G simulator and added support for multiple RATs. The source code of the 5G multi-RAT simulator has been released under MIT License and is available online~\cite{code}. In our simulator, packets are the fundamental objects, and we have implemented physical and MAC layer protocol stacks for 5G-NR and Wi-Fi RATs. Support for higher layer protocol stacks such as RRC and NGAP is not added at present. However, we have implemented a centralized controller that manages these RATs. 
    
    \begin{figure}
        \centering
       \includegraphics[scale=0.58,trim = {1.5cm 0.5cm 1cm 0.8cm},clip]{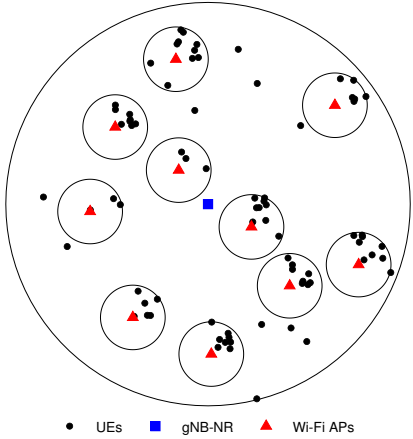}
        \caption{Simulation Scenario}
        \label{fig:sim_scenario}
    \end{figure}
    
    We consider a single-cell scenario with $250$ m radius as shown in Fig.~\ref{fig:sim_scenario}. The network model comprises of a 5G NR cell and multiple Wi-Fi Basic Service Areas (BSAs). The cell consists of a gNB-NR entity located at the center. The 5G-Flow controller and the MRN OF switch can either be co-located with the gNB-NR entity or can be cloud-based. Multiple Wi-Fi APs provide radio coverage (BSA) inside the gNB-NR cell and are distributed uniformly at random. A UE in the network is assumed to have two radio interfaces: 5G-NR and Wi-Fi. We also assume that $80\%$ of users are connected to both the RATs while the remaining users are connected to only gNB-NR entity. The users, in our simulation model, are assumed to be stationary.

    \begin{table}[!b]
        \centering
        \caption{Network Model}
        \label{tab:param5Gflow}
        \renewcommand{\arraystretch}{0.9}
        \begin{tabular}{p{3.5cm}p{3.5cm}}
            \hline \hline
            \textbf{Parameters} & \textbf{Values}   \\ \hline 
            Number of Wi-Fi APs & 10 \\ 
            Number of Users & 80 \\ 
            Packet Payload & 1000 bytes \\
            TCP Header & 60 bytes\\ \hline 
            \multicolumn{2}{c}{5G-NR Network Model} \\ \hline 
            Carrier Frequency    & $1.9$ GHz (TDD) \\
            5G Numerology & $1$  \\
            Bandwidth   & $60$ MHz (162 PRBs, 2 slots per sub-frame)\\ 
            UE/gNB Transmit Power & $43/23$ dBm     \\
            UE/gNB Antenna Gain  & $2$/$15$ dBi      \\ 
            UE/gNB Antenna Height  & $1.5$/$25$ m    \\ 
            UE/gNB Noise Figure	  & $7$/$10$ dB		\\ 
            \hline 
            \multicolumn{2}{c}{Wi-Fi Network Model} \\ \hline 
            Bandwidth   & $20$ MHz         \\   
            Operating Frequency & $2.4$ GHz \\ 
            Coverage radius & $40$ m \\ 
            UE/AP Transmit Power & $20/15$ dBm             \\ 
            AP Antenna Gain  & $4$ dBi      \\ 
            AP Antenna Height  & $10$ m    \\ 
            MPDU & $1500$ bytes \\ \hline \hline
          \end{tabular}
        \end{table}

    For 5G-NR, we use 3GPP Urban Macro (UMa) path loss model, whereas Urban Micro (UMi) path loss model for Wi-Fi ~\cite{38.901,802.11model}. We also consider log-normal shadow fading for 5G-NR and Wi-Fi RATs and their standard deviations are as per the UMa and UMi path loss model. The Wi-Fi network is based on IEEE 802.11n~\cite{802.11}. Other important parameters used in our simulator are given in Table~\ref{tab:param5Gflow}.
    
    We perform Monte Carlo simulations for 1 second and average the results over 50 deployment scenarios. We analyze the performance of average TCP throughput and average packet delay for the 5G-Flow network and compare them with the current 3GPP 5G network. In our simulation model, the TCP throughput is the sum of TCP throughput of all the users in a cell. We define packet delay as the total time it takes for a packet to reach its destination node from the source node.

   \subsection{Downlink dataflow management for UEs with different service types} 
    As discussed before, data flow management involves selecting an optimal RAT (from the available RATs) for each session of a UE based on the RAN-level information.  To analyze the performance of 5G-Flow RAN with respect to RAT selection, we consider four types of data services offered in the network in the order of the priority, i)  mission-critical streaming, ii) real-time streaming, iii) non-real-time streaming and iv) best-effort. For simplicity, we assume that a UE requests only one type of service for the entire duration of simulation. We assume Poisson traffic with the packet arrival rate of $500$ packets/sec. The packet size is fixed at 1000 bytes (payload). Therefore, the downlink bit-rate is $4$ Mbps for all the users (irrespective of the service requested), and TCP protocol is used as transport protocol. Since we aim to analyze the downlink performance, we assume that the entire available bandwidth is used by the downlink traffic and do not consider the bandwidth used by TCP-ACK packets (in uplink). We evaluate the uplink performance of the proposed network in the scenario discussed next. We assume $80$ users in our simulation model. We consider five different cases based on the service type requested by a user and these are explained in Table~\ref{tab:cases}. The 5G-NR radio resource scheduler uses a priority scheduling algorithm to give better service to higher priority users. Wi-Fi network uses CSMA/CA to schedule the users and does not distinguish between user service priorities.
    
     \begin{table}
        \centering
        \caption{Simulation Cases for 80 Users in a Cell}
        \begin{tabular}{ccccc}
            \hline
            Cases & \multicolumn{4}{c}{Service Type}  \\
            & 1 & 2 & 3 & 4 \\ \hline
            Case a & 0 & 0 & 40 & 40 \\
            Case b & 10 & 10 & 30 & 30 \\
            Case c & 20 & 20 & 20 & 20 \\
            Case d & 30 & 30 & 10 & 10 \\
            Case e & 40 & 40 & 0 & 0 \\ 
            \hline
        \end{tabular}
        \label{tab:cases}
    \end{table}

   For the purpose of performance evaluation, we have designed a threshold-based heuristic algorithm for RAT selection given in Algorithm $1$.. The algorithm considers the following metrics. The load conditions at RATs (Wi-Fi and gNB-NR) and 5G-NR radio channel quality are the most critical metrics to select a RAT for UE efficiently. In addition, the RAT selection algorithm also takes into account the type (priority) of service being used. Due to the small coverage area of Wi-Fi AP, received signal strength for UEs do not vary much. Hence, the channel condition for a UE under Wi-Fi network is always considered good. We assign different weights for the different metrics and calculate the value of $T_i$ (for every UE) based on the following equation. We then select the RAT for a UE based on the predetermined threshold value. 
        \begin{align}
            \centering
            \label{thresh_eq}
            T_i(l_{g},&l_{w},ch_{g,i},s_i) = \alpha\cdot l_{g} + \beta\cdot l_{w} + \gamma\cdot ch_{g} + \delta\cdot s_i \\
       \text{where,     }& \nonumber\\
       l_{g} =& \text{ Load at gNB} \nonumber\\
       l_{w} =& \text{ Load at Wi-Fi AP to which the user is connected} \nonumber\\
       ch_{g,i} =& \text{ Channel condition for user $i$ under 5G-NR network} \nonumber\\
       s_i =& \text{ Service type of user $i$} \nonumber
    \end{align}
   Let $L=\{1,2,3\}$ denote the set of values representing low, medium and high load respectively at a particular RAT. The load at gNB ($l_{g}$) and Wi-Fi ($l_{w}$) take its values from $L$. The channel condition experienced by user $i$ from gNB is represented by $ch_{g,i}\in\{0,1\}$, where $0$ represents good channel and $1$ represents bad channel. We distinguish between good and bad channel based on a threshold value of received Signal-to-Noise Ratio (SNR), which has been fixed at $6$ dB in the simulation model. The service type of user $i$ is represented by $s_i \in \{1,2,3,4\}$, where $1$ represents the highest priority service while $4$ represents the service type with the least priority. The coefficients in Eq.~\ref{thresh_eq} ($\alpha, \beta, \gamma$ and $\delta$) are the weights assigned to the metrics in the equation. The weights can be modulated based on the impact of a certain metric on system performance. The weights considered in our algorithm are given in Algorithm $1$. We assign highest importance to $ch_{g,i}$ because bad channel leads to resource wastage and poor performance. Since L1/L2 efficiency of gNB-NR RAT is better than that of Wi-Fi, we give higher importance to the load at gNB. After the evaluation of $T_i$, RAT selected for a user $i$ ($R_i$) is determined based on the threshold value $T^{'}$. Since, we aim to balance the load on the available RATs, $T^{'}$ is set to be the mean of all possible values of $T_i$.
    \begin{align*}
         & \quad \quad R_i = \left\{\begin{matrix}
            1, & T_i > T^{'} \quad(1 \text{ represents Wi-Fi} )\\ 
            0, & T_i \leq T^{'} \quad ( 0 \text{ represents gNB})
            \end{matrix}\right.
    \end{align*}
    
   Since the 5GC is unaware of the RAN level information, the RAT selection policy for standard 3GPP 5G network only considers the service priority. For the performance evaluation of 5G network, we consider that a user with service priority $1$ or $2$, is always scheduled at gNB and with service priority $3$ or $4$ is always scheduled at Wi-Fi (irrespective of load or channel condition).

    \begin{algorithm}
	    \DontPrintSemicolon
	     \SetAlgoLined
		\caption{RAT Selection Algorithm for Downlink}
		\KwInput{\;
		$C_{g} = C_0$ \tcp*{gNB channel capacity} 
		$C_{w} = W_0$ \tcp*{Wi-Fi channel capacity}
		$l_{g} = 1, l_{w} = 1$  \tcp*{Starting at low load}}
		$\alpha = 30, \beta = 10, \gamma = 50, \delta = 25$ \;
		$T^{'}= 170$ \;
		\tcp{$N=$ Number of UEs associated with both RATs}
		
		\For{$i$ from $1$ to $N$}
		{
		Evaluate $ch_{g,i}$ for user $i$\;
		Calculate $T_i(l_{g},l_{w},ch_{g,i},s_i)$ based on Eq. (1) \;
		\eIf{$T_i > T^{'}$}
		{
        $R_i = 1$ \;
        Decrease $C_{w}$ \;
        \If{$C_{w}$ = medium $||$ $C_{w}$ = low}{
        Update $l_{w}$}}
        {
         $R_i = 0$ \;
        Decrease $C_{g}$ \;
        \If{$C_{g}$ = medium $||$ $C_{g}$ = low}{
        Update $l_{g}$}}}
        \KwOutput{$R_i , {\forall} i \in \{1,2,\ldots,N\}$ }
	\end{algorithm}

\begin{figure}
        \begin{subfigure}{0.49\textwidth}
            \centering
            \includegraphics[height=4.2cm,width=4.4cm]{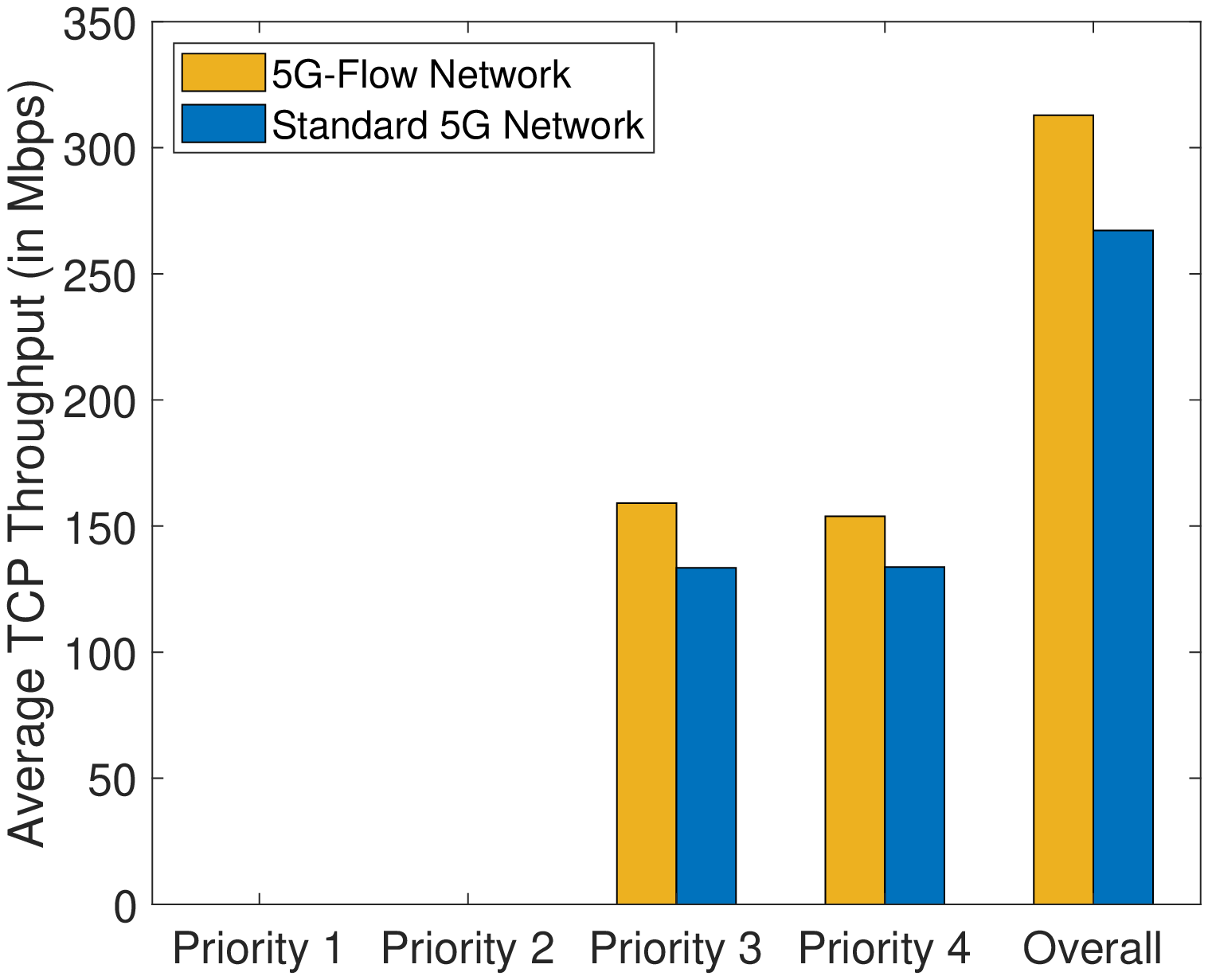}%
            \includegraphics[height=4.2cm,width=4.4cm]{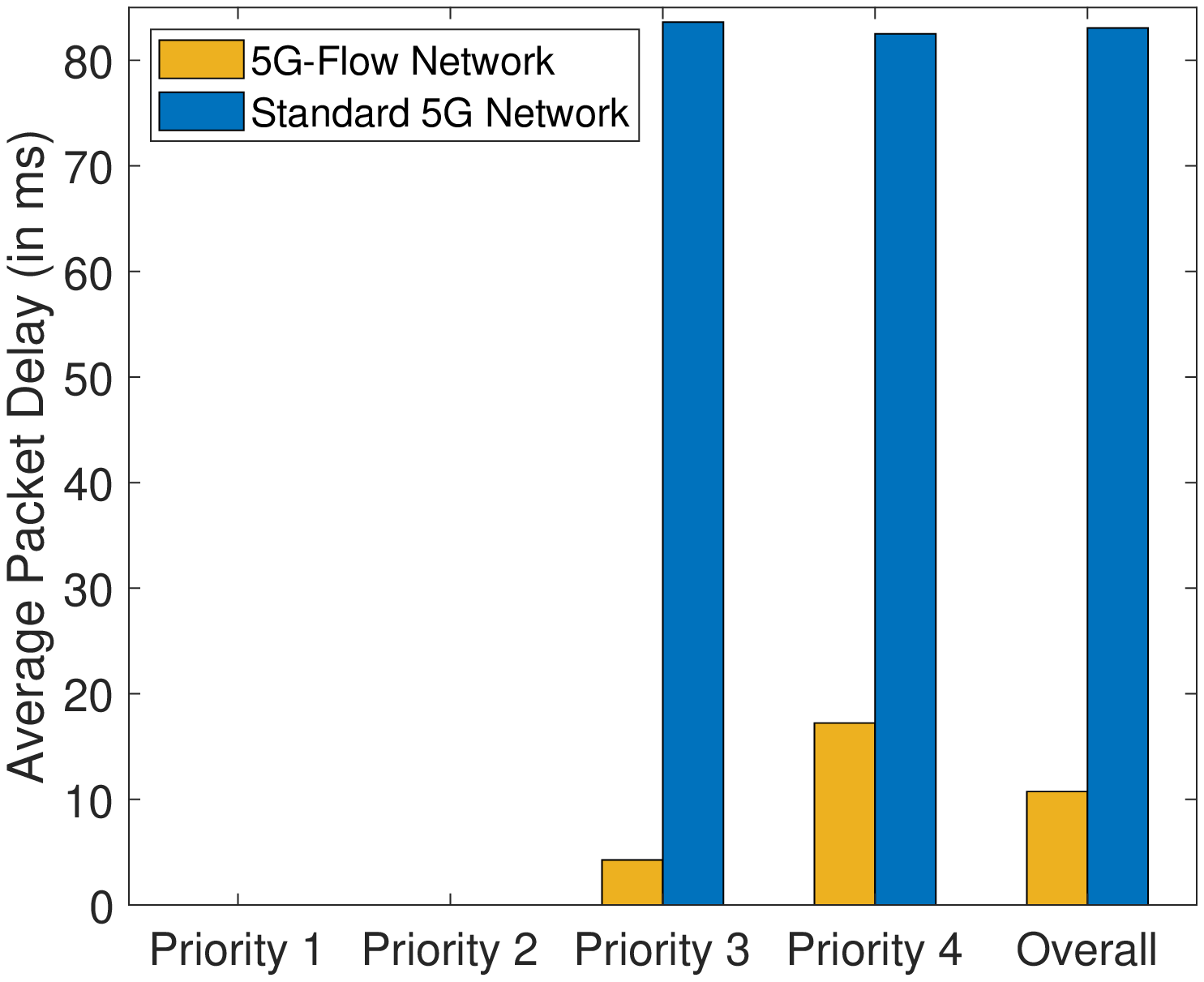}
            \vspace{-0.6cm} 
            \caption{Case a}
            \label{fig:0_0_40_40}
        \end{subfigure}
        \begin{subfigure}{0.49\textwidth}
            \centering
            \includegraphics[height=4.2cm,width=4.4cm]{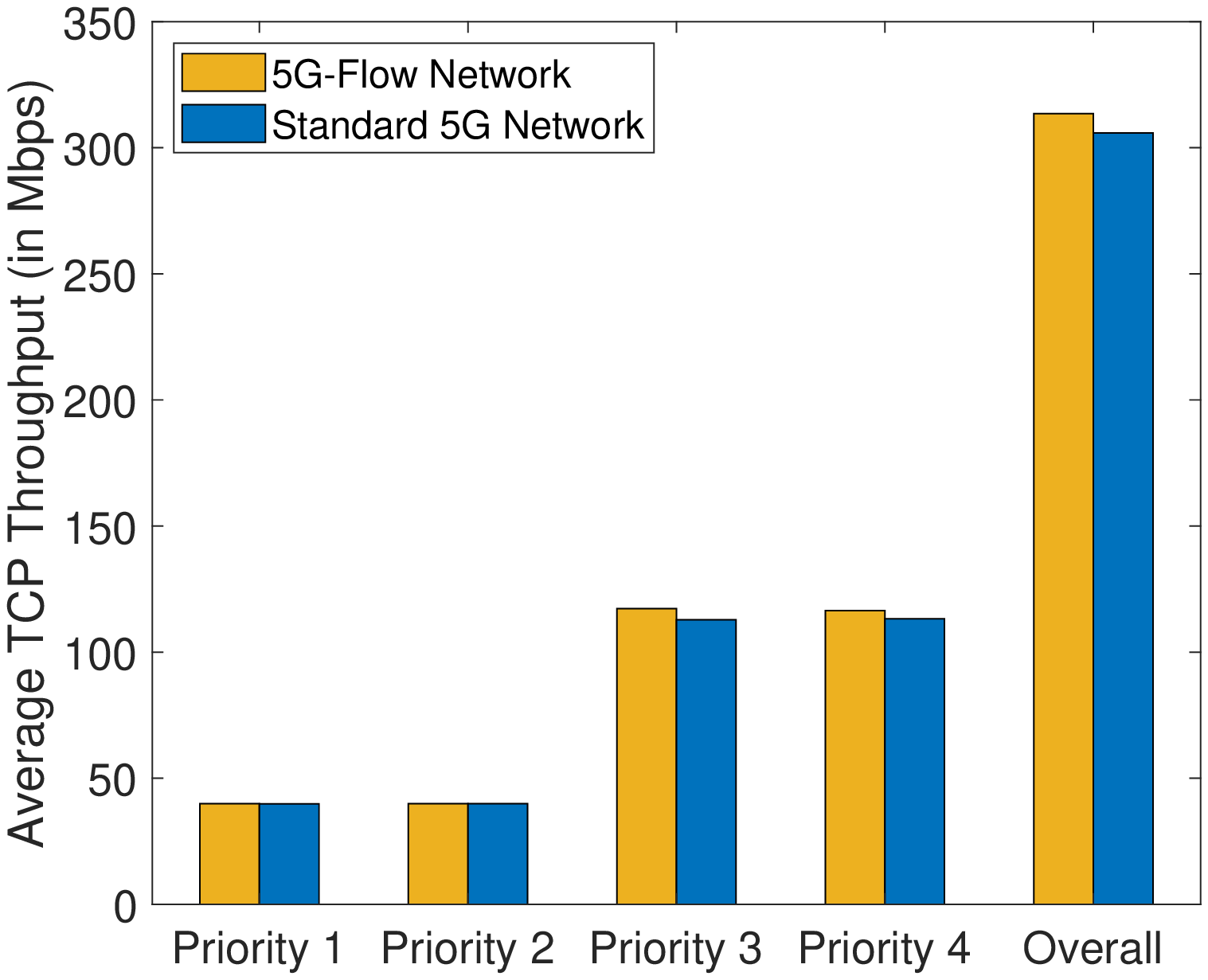}%
             \hfill
            \includegraphics[height=4.2cm,width=4.4cm]{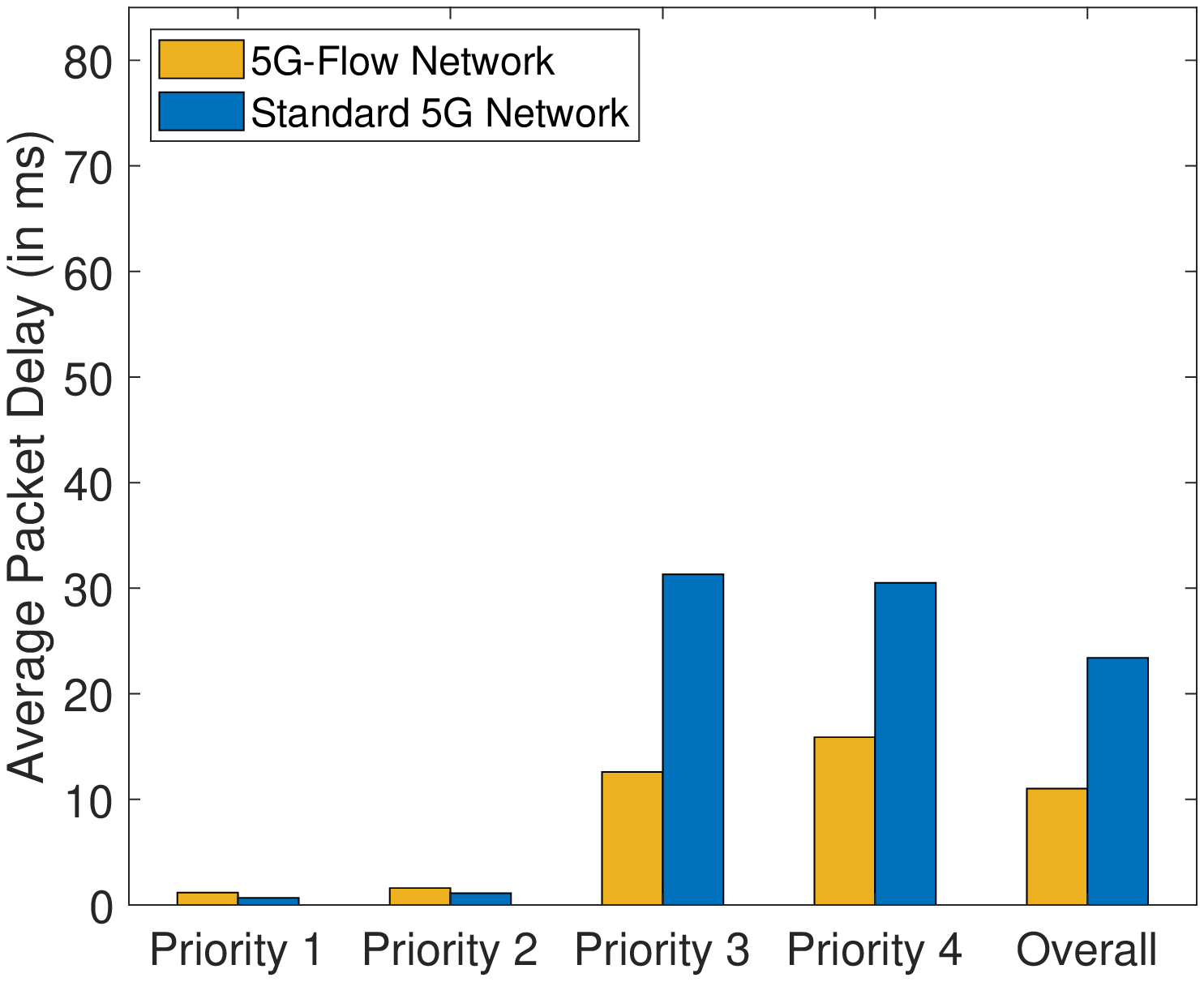}
            \vspace{-0.6cm} 
            \caption{Case b}
            \label{fig:10_10_30_30}
        \end{subfigure}
        \begin{subfigure}{0.49\textwidth}
            \centering
            \includegraphics[height=4.2cm,width=4.4cm]{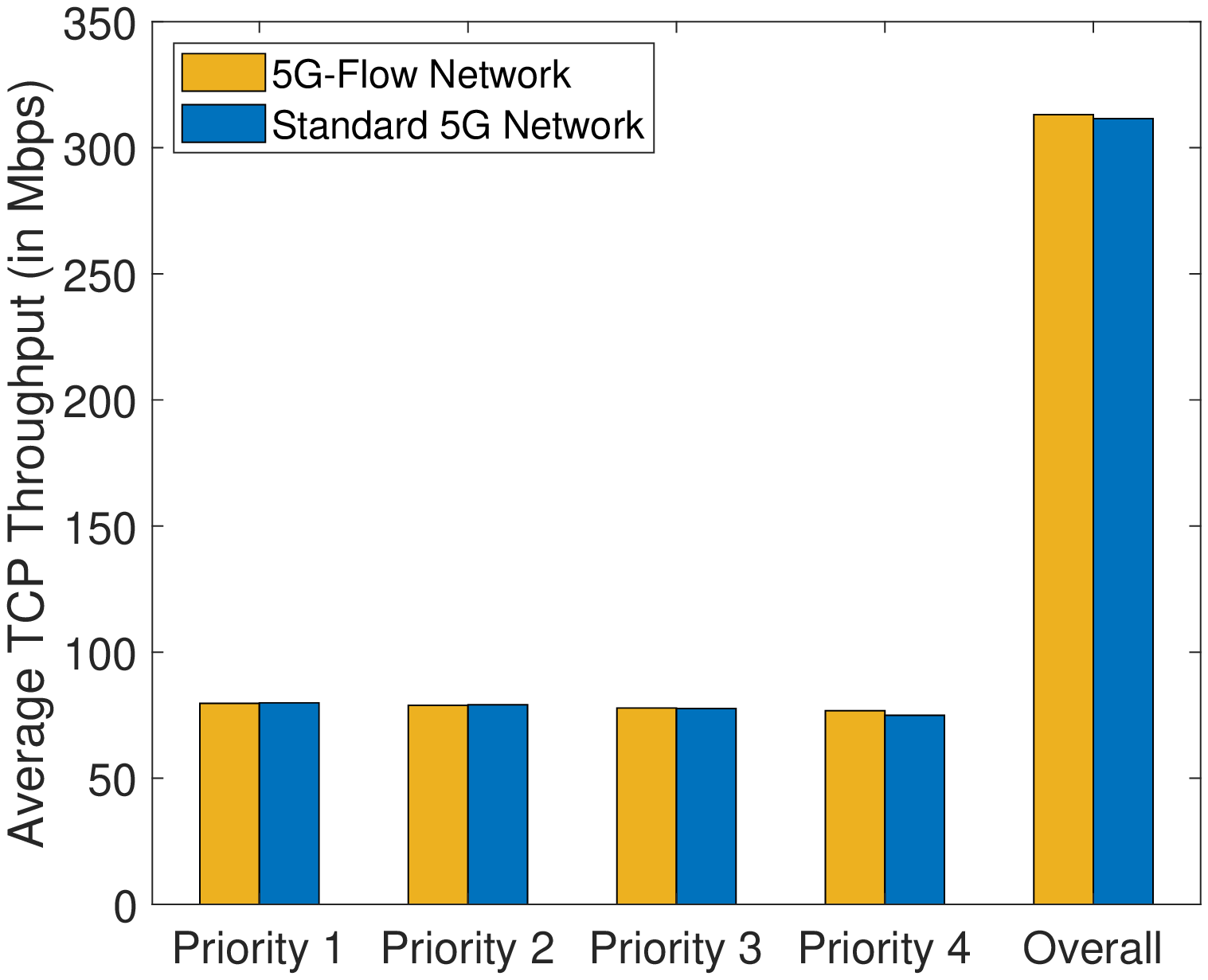}%
             \hfill
            \includegraphics[height=4.2cm,width=4.4cm]{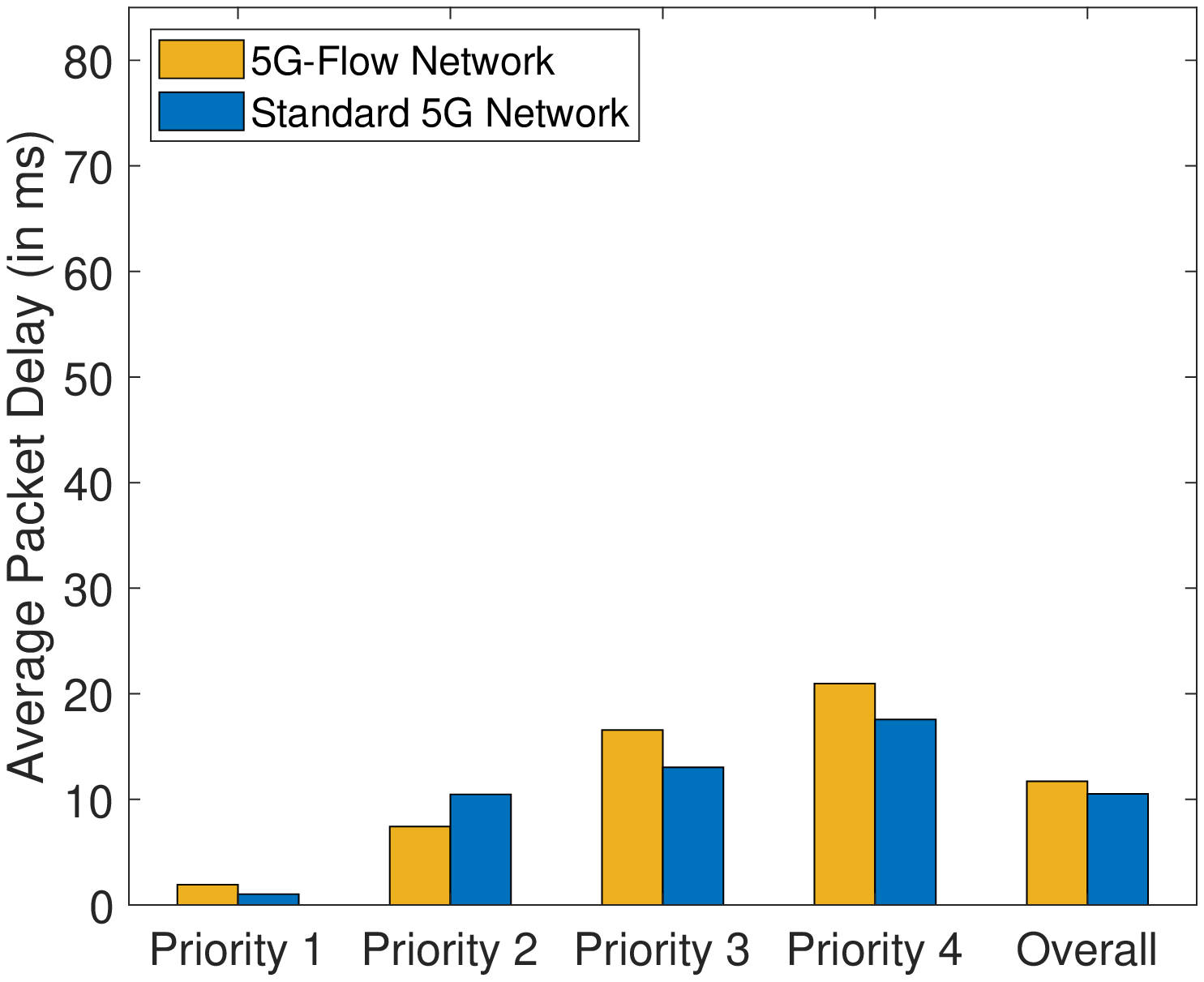}
            \vspace{-0.6cm} 
            \caption{Case c}
            \label{fig:20_20_20_20}
        \end{subfigure}
        \begin{subfigure}{0.49\textwidth}
        \centering
            \includegraphics[height=4.2cm,width=4.4cm]{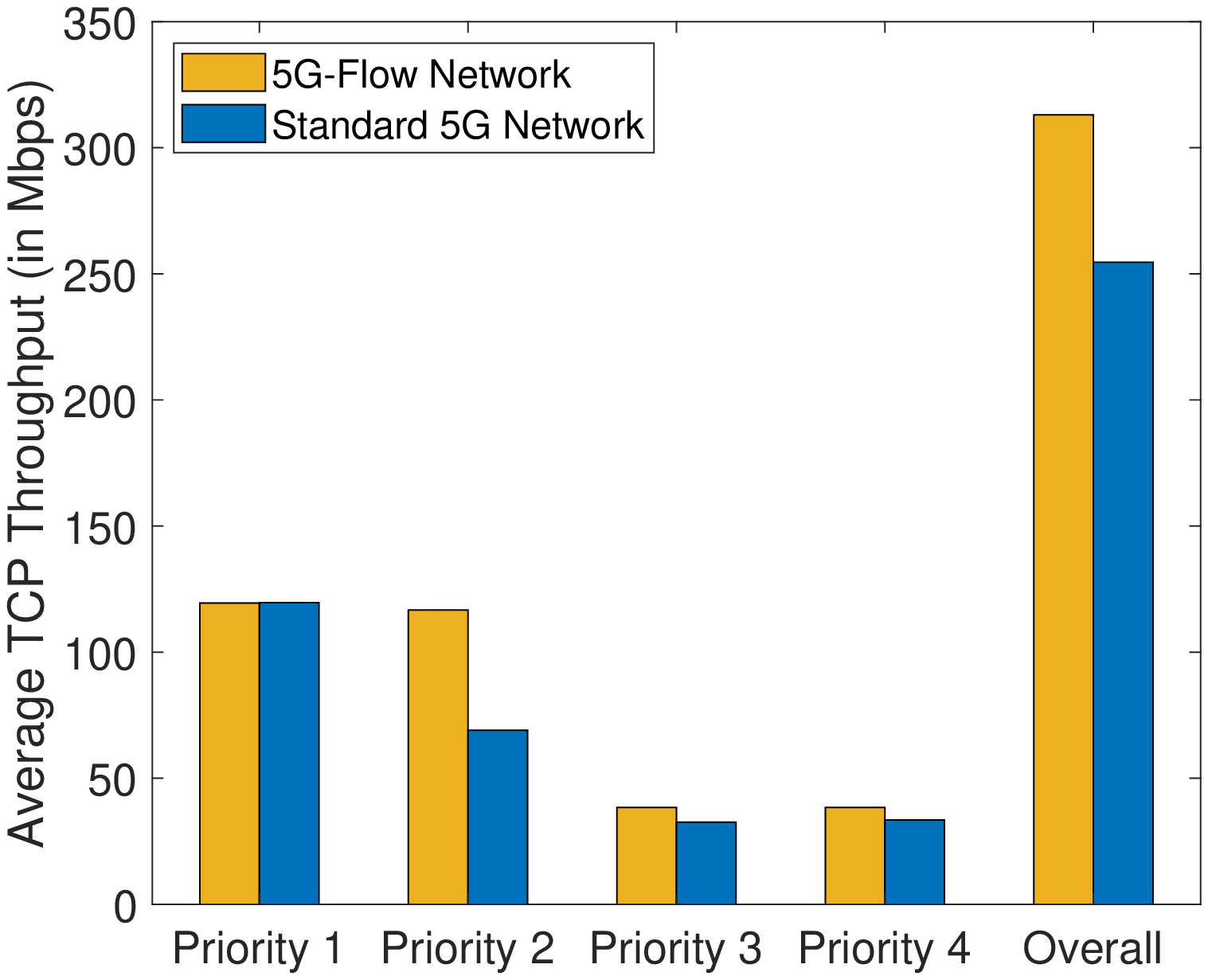}%
            \hfill
           \includegraphics[height=4.2cm,width =4.4cm]{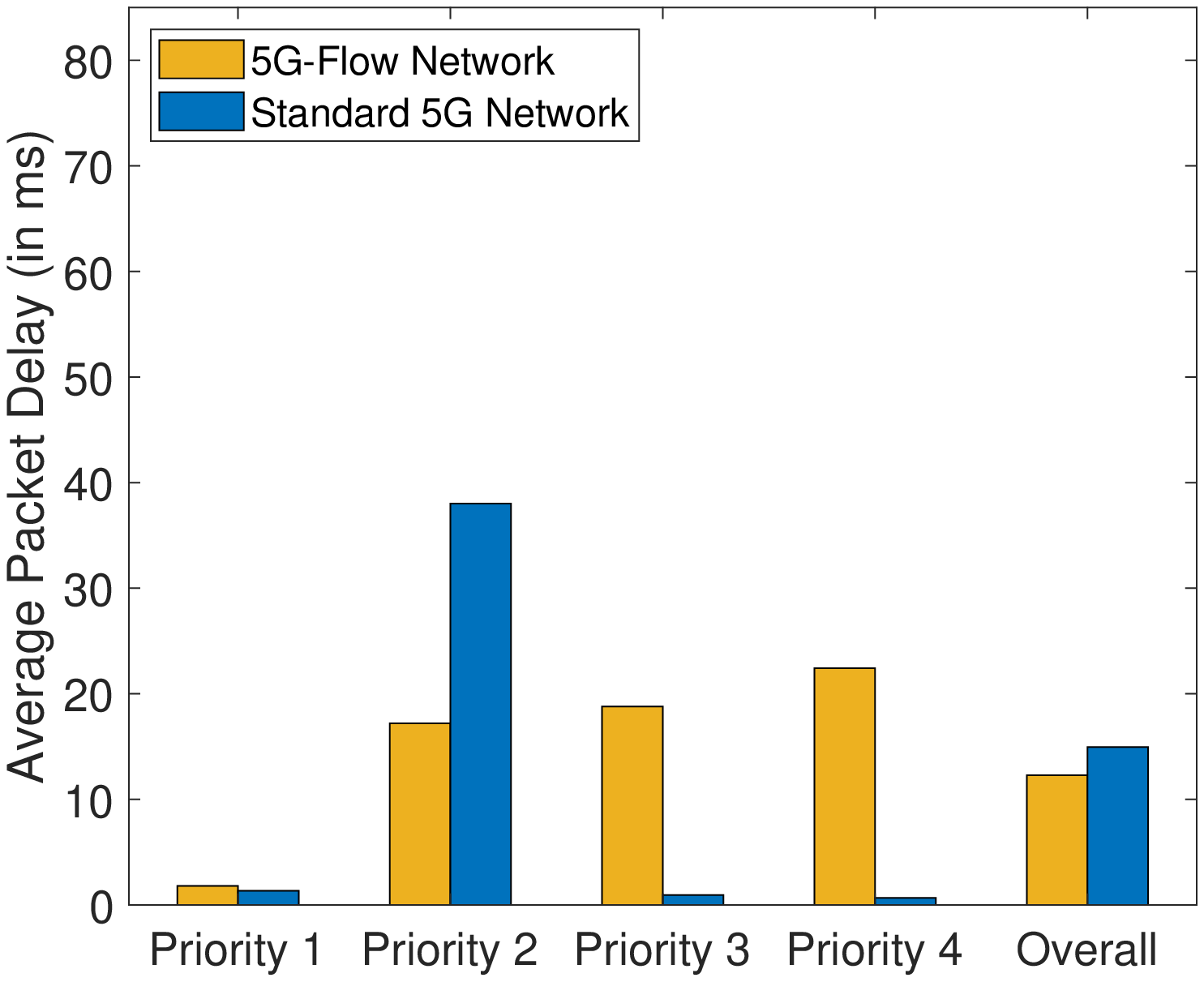}
           \vspace{-0.6cm} 
           \caption{Case d}
            \label{fig:30_30_10_10}
        \end{subfigure}
        \begin{subfigure}{0.49\textwidth}
            \centering
            \includegraphics[height=4.2cm,width=4.4cm]{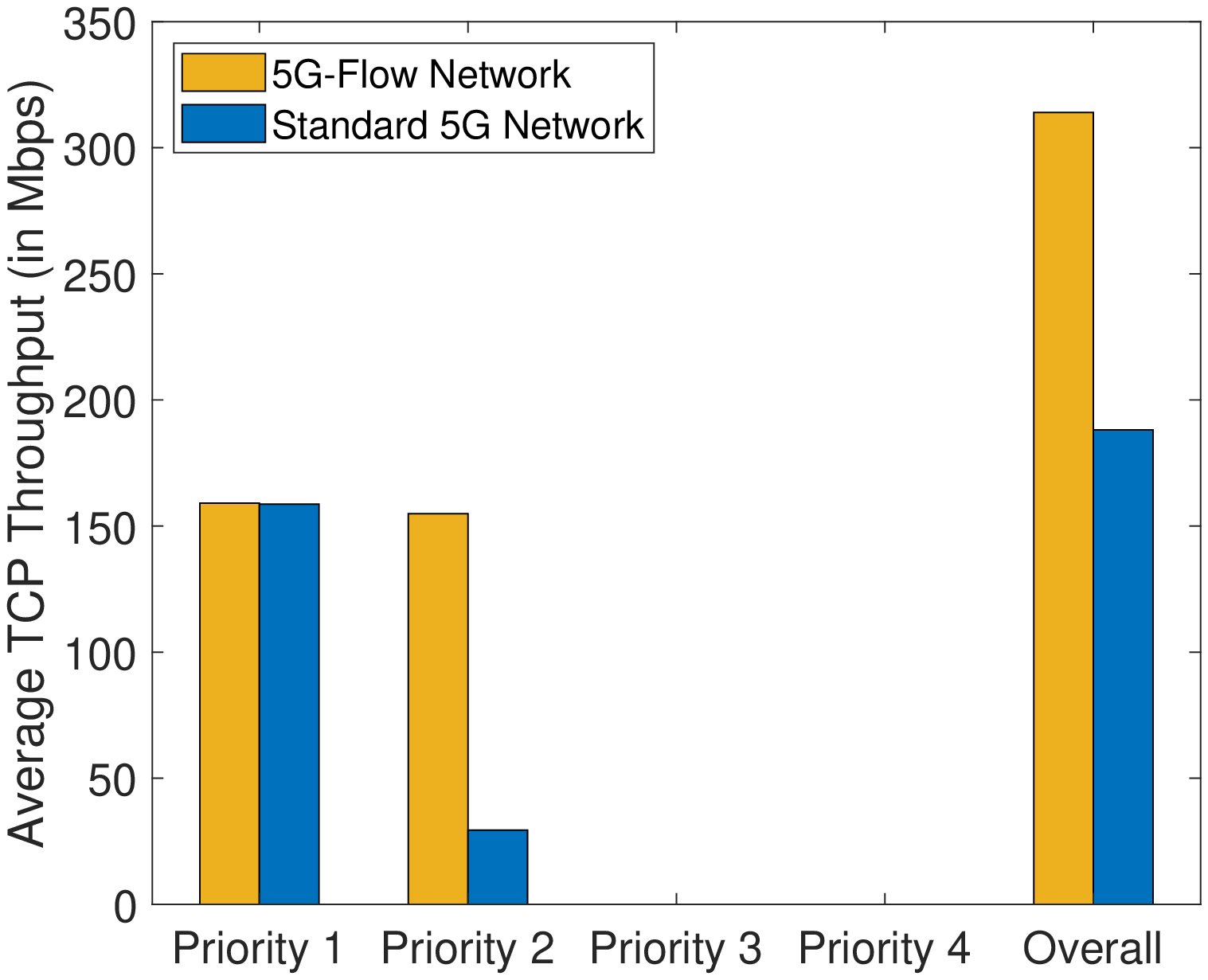}%
             \hfill
            \includegraphics[height=4.2cm,width=4.4cm]{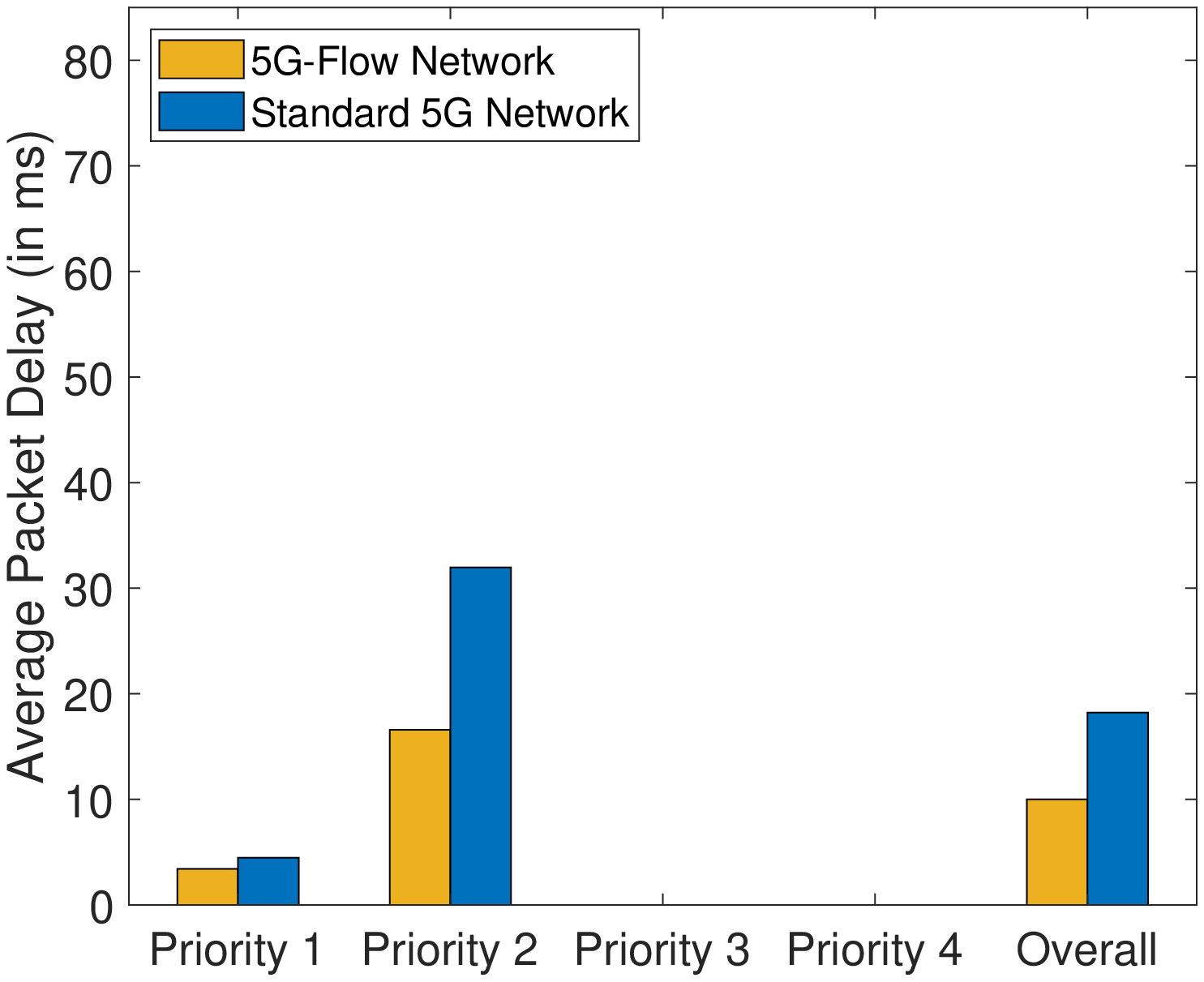}
            \vspace{-0.6cm} 
            \caption{Case e}
            \label{fig:40_40_0_0}
        \end{subfigure}
        \caption{Downlink Performance of 5G-Flow network vis-a-vis 3GPP 5G network based on simulation cases considered in Table~\ref{tab:cases}.}
        \label{fig:rat_selec}
        
    \end{figure}

     \textit{Results}: Fig.~\ref{fig:rat_selec} shows the average TCP throughput and packet delay for various simulation cases considered in Table~\ref{tab:cases}. We first analyse the graphs in Fig.~\ref{fig:0_0_40_40} and Fig.~\ref{fig:40_40_0_0}, for extremely skewed service type distribution among users. Since the 5G-Flow network considers load on RATs and channel condition, RAT selection is done in an efficient way. Hence, the throughput and delay performance for 5G-Flow network are significantly better. Since all the users in standard 5G network are scheduled either on Wi-Fi RAT (in Fig.~\ref{fig:0_0_40_40}) or on gNB-NR RAT (in Fig.~\ref{fig:40_40_0_0}), the respective RAT is overloaded and hence the overall performance is poor. This also leads to poor utilization of the other RAT.      
    
    Fig.~\ref{fig:10_10_30_30} and Fig.~\ref{fig:30_30_10_10} illustrate average throughput and packet delay results for service type distributions which are relatively less skewed than the cases discussed before. In Fig.~\ref{fig:10_10_30_30}, the average packet delay under standard 5G network is quite high for users with service priority $3$ and $4$ as all the $60$ users (with these service priorities) are scheduled at Wi-Fi. gNB under this case, experiences low load and hence the average packet delay is quite low for users with service priority $1$ and $2$. Similarly, in Fig.~\ref{fig:30_30_10_10}, the delay performance of users with service priority $2$ suffers as the gNB experiences traffic load from $60$ users and it prioritises the users with priority $1$. The performance of evenly distributed user service priority is shown in Fig~\ref{fig:20_20_20_20}. The performances of 5G-Flow network and standard 5G network are comparable as the 5G network follows a RAT selection policy which is best suited for evenly distributed user service priority. 
    
   In general, the standard 5G network is unable to efficiently use the available radio resources in a multi-RAT scenario. The performance of the 5G-Flow network significantly improves as it considers RAN-level information while performing RAT selection. 
    
    \begin{figure}
        \centering
        \includegraphics[height=4.1cm]{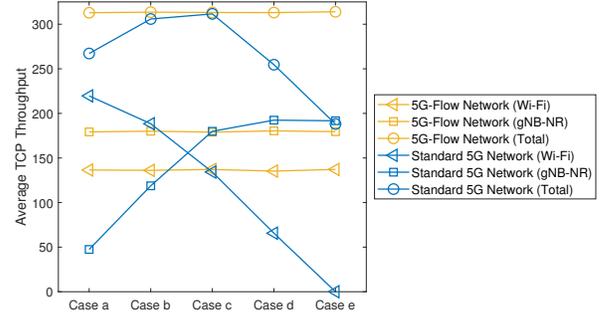}
        \caption{Load balancing across RATs}
        \label{fig:load_balancing}
    \end{figure}
    
    \textit{Load Balancing}: In Fig.~\ref{fig:load_balancing}, the graph shows how the traffic is split across various RATs for simulation cases presented in Table~\ref{tab:cases}. The 5G network in unable to balance the traffic across RATs and hence the overall throughput suffers. On the other hand, 5G-Flow network observes a consistent performance as it balances the load across RATs.

    \subsection{Uplink Dataflow management}
    The proposed 5G-Flow network enables flexibility in managing uplink and downlink traffic of every UE independently. Moreover, our architecture decouples uplink and downlink traffic management, i.e., RAT selection for a UE's uplink and downlink traffic is determined independently. Due to this decoupling, we expect that the performance of the overall network and, in particular, of Wi-Fi network will improve when the uplink users are fewer in a Wi-Fi network. This is expected as fewer users will contend to gain access to the channel in a CSMA/CA system. 

    In the existing 3GPP 5G network, ATSSS feature allows for independent decisions on uplink and downlink traffic distribution. UE selects RAT for uplink transmission based on channel quality that it is observing on multiple RATs and ATSSS rules. UE does not know the load on each RAT to decide the best possible RAT for uplink.  For instance, let us assume that a UE observes a good channel gain from Wi-Fi AP as well as gNB-NR, so it selects gNB-NR interface for uplink. gNB-NR may be heavily loaded, and Wi-Fi may be lightly loaded, leading to poor uplink performance (which could have been better on Wi-Fi). Therefore, without loss of generality, we assume that UEs select the same RAT for uplink and downlink i.e., uplink and downlink are essentially coupled for 5G network.
    
    \begin{algorithm}
        \SetAlgoLined
        \DontPrintSemicolon
		\caption{RAT Selection Algorithm for Uplink}
		\KwInput{ \;
		$C_{g} = C_0$ \tcp*{gNB-NR Uplink channel capacity}
		\tcc{$W_0 = $ Maximum number of users allowed to connect with a Wi-Fi AP in uplink}}
		$A =$ List of sorted Wi-Fi APs with respect to number of connected UEs\;
		
			\For {each $i$ in $A$}{
    			$N_i$ = Number of users connected to Wi-Fi AP $i$\;
    			$K = \min (N_i - W_0, C_{g})$\;
    			Move $K$ users (closest to gNB) from Wi-Fi to gNB-NR for uplink data transmission\;
    			Update $C_{g}$
    			}
                
	\end{algorithm}

    In our simulation scenario, we use TDD duplex scheme to allocate resources for uplink and downlink. The transmission periodicity for the TDD radio frame is $5$ ms, i.e., the frame repeats after $10$ slots (for 30 khz sub-carrier spacing). TDD frame configuration is \{D,D,D,S,U,U,U,U,U\}, where D, U, and S represent downlink, uplink and special slot respectively~\cite{38.331}. The S slot consists of downlink symbols except for the last symbol, which is used as a switching symbol. We consider 80 users each having $3$ Mbps downlink and $1$ Mbps uplink data rate requirement. We consider the Poisson traffic model. We do not consider service priority in this simulation scenario. Therefore, we use a round-robin scheduler for 5G-NR to allocate radio resources to the users.
    
    \begin{figure}
        \centering
        \includegraphics[width=4.4cm,height=4.01cm]{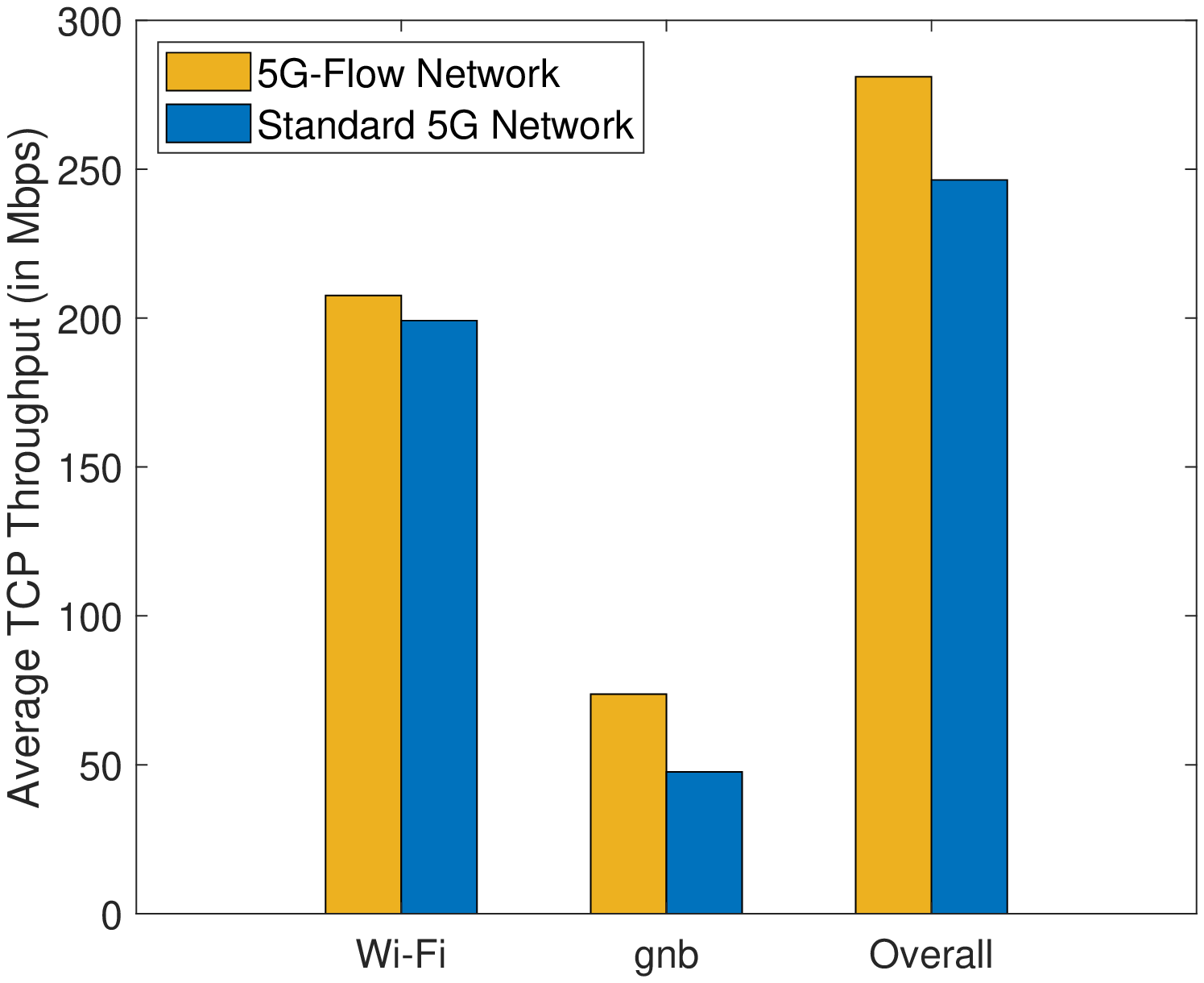}%
        \hfill
        \includegraphics[width=4.4cm,height=4.01cm]{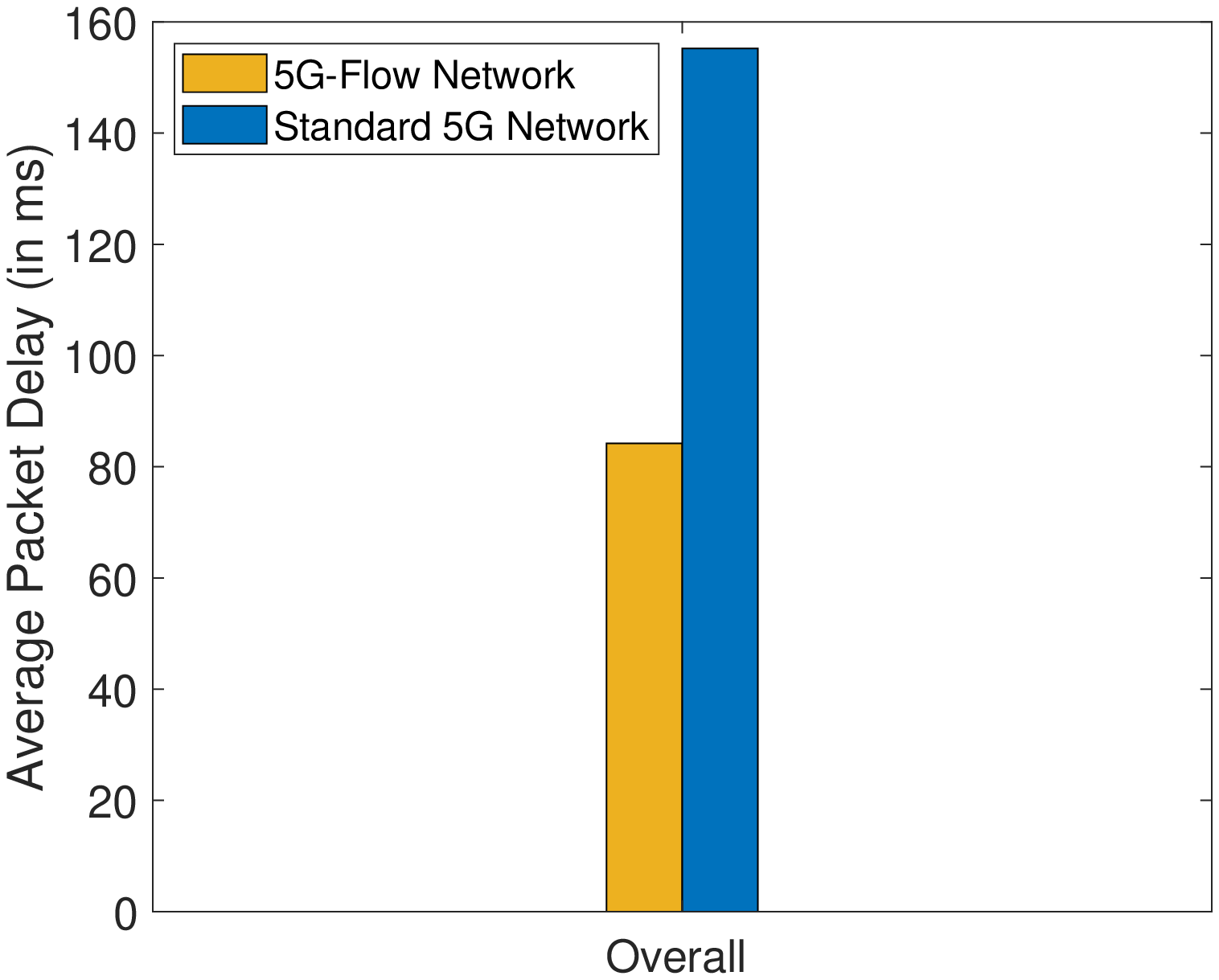}
        \caption{Performance comparison between 5G-Flow network and Standard 5G network when Uplink and Downlink data flows are decoupled.}
        \label{fig:uldl}
    \end{figure}
    
    For those users that are dual-connected in the 5G-Flow network, downlink data is scheduled on Wi-Fi. However, we schedule the uplink data to gNB-NR interface for users who are experiencing good channel gain from the gNB. The RAT selection algorithm for uplink users is given in Algorithm 2.  

    \textit{Results}: Fig.~\ref{fig:uldl} shows that Wi-Fi as well as the overall TCP throughput improves for the 5G-Flow network. It is important to note that throughput improvement is observed after the load on Wi-Fi is reduced (under 5G-Flow network) as we have offloaded the uplink of some users to gNB-NR. The throughput gain can be attributed to reduced number of users contending to gain channel access in 5G-Flow network. For the same reason, the delay performance has also significantly improved.

   \section{Applications}
  In this section, we present some important applications of 5G-Flow RAN architecture below. 
   \subsection{Connecting to the Internet Directly via RAN}
\label{sec:internet}
A UE in the existing 3GPP cellular network can access Internet solely through the CN. It does not have the flexibility to connect to the Internet directly from the RAN bypassing the CN. This feature can be beneficial in those areas where users are mostly stationary, and therefore a data tunnel through CN may not be required. However, the advantages of the cellular stack with an efficient L2/L1 layer can still be exploited. 

We now discuss the procedure to access the Internet directly bypassing the CN. A UE sends a packet to the IP interface of the UE OF switch and a table-miss is observed. On table-miss, the UE OF switch forwards the packet to the 5G-Flow controller via \textit{OF Packet-in} message. Depending on the QoS requirement, the controller decides whether a dedicated DRB needs to be created. In case a dedicated DRB is required, the controller sends an \textit{OF-Config} message to gNB-NR interface to create a logical port, which is translated by RRC layer and a DRB is created using \textit{RRC Reconfiguration} messages. An OF-Config message is also sent to the UE so that it can create a logical port on the NR interface and map the newly established DRB to the logical port. The 5G-Flow controller adds the appropriate flow entries at UE and MRN OF switch so that the UE can access the Internet.
    
    \subsection{Simpler Mechanism for 5G Non-standalone Deployment}
\begin{figure}
        \centering
        \subcaptionbox{}{\includegraphics[scale=0.6,trim={0 13cm 20cm 0},clip]{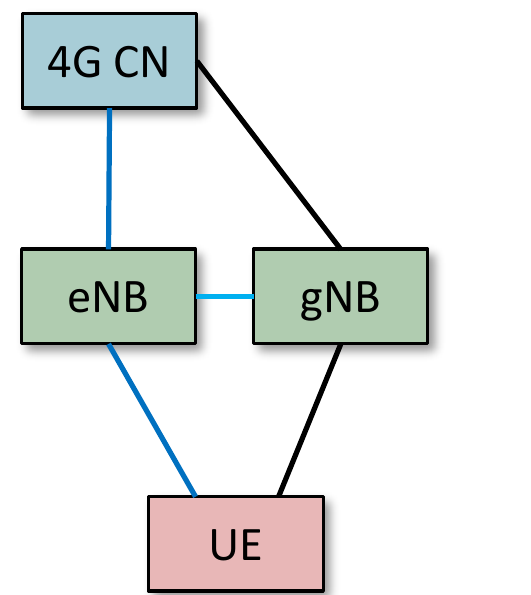}}%
        \subcaptionbox{}{\includegraphics[scale=0.6,trim={0 13cm 18cm 0},clip]{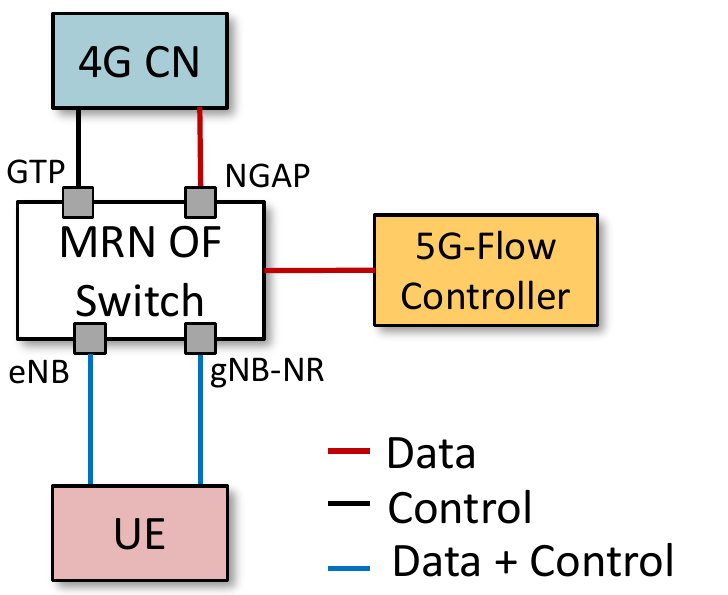}}
        \caption{Implementation of non-standalone 5G architecture proposed by (a) 3GPP, viz-a-viz (b) 5G-Flow Network.}
        \label{fig:nsa}
    \end{figure}

    It appears that 5G deployments will be carried out in phases, and the early adopters of 5G are most likely to choose a non-standalone deployment. This method involves deployment of eNB and gNB in RAN, which communicate with the 4G CN as shown in Fig.~\ref{fig:nsa}. A UE should support dual connectivity to both eNB and gNB to avail 5G services. Since gNB does not communicate with 4G CN, it is modified to communicate with 4G CN and referred to as en-gNB.  An eNB acts as a master node while en-gNB acts as a secondary node, and they communicate via X2 interface. All signaling exchange between UE and the network happens via eNB, while gNB is only used for data transfer. The non-standalone method is a much faster way to deploy 5G as it leverages the existing 4G infrastructure. However, 5G RAN capability can not be exploited entirely in the current architecture. 
    
    5G-Flow architecture enables non-standalone deployment in a much simpler manner. Due to the complete decoupling between RAN and CN connectivity, it allows a UE to use 5G NR interface to connect to 4G CN and receive mobile data services without necessarily using the dual connectivity. At the same time, 5G-Flow also enables a UE to use dual connectivity, whenever required.   Our proposed 5G-Flow RAN architecture requires minimal changes that can be easily adopted in the 4G framework. It allows a 4G/5G capable UE to communicate with 4G CN via eNB as well as gNB. The control and data signaling can be transported via both eNB and gNB, thereby enhancing the RAN efficiency. Additionally, all the benefits of 5G-Flow RAN can also be exercised.

\section{Conclusions and Discussions}
Unified multi-RAT systems are indispensable in the next-generation networks. Although 3GPP 5G architecture supports multi-RAT integration at 5GC, there are several gaps in the existing 5G architecture. To address this, we re-architect the 3GPP 5G RAN to realize an integrated, software-defined multi-RAT RAN using OpenFlow Protocol. To realize the 5G-Flow RAN, we have suggested minimal software-based changes in the 3GPP 5G architecture which can be easily implemented. Moreover, we do not propose any changes in the protocol used between UE and gNB, UE and 5GC, and gNB and 5GC. The proposed architecture presents significant advantages over existing 3GPP 5G architecture such as i) simplified multi-RAT integration, ii) efficient dataflow management at RAN, iii) flexibility to connect to any CN or Internet directly via 4G/5G/Wi-Fi based RAN, and iv) simpler implementation of the non-standalone architecture. The performance evaluation of our architecture (using the simulator developed by us) shows promising gains over the 3GPP 5G network performance. 

The current 3GPP 5G architecture is designed in a manner that drives user traffic through CN even when the user is stationary. In this paper, we present a method wherein a user can bypass CN and directly access Internet. This feature can be beneficial in effectively managing future networks that aim to serve billions of connected devices, of which a large fraction belongs to stationary IoT devices. Complete decoupling of radio and CN protocol stack in 5G-Flow RAN allows for greater flexibility to develop radio technology independent of the CN being used. Future networks can exploit this feature to facilitate faster development cycles for newer RATs.

The development of an open protocol for analyzing and translating NGAP messages for the 5G-Flow controller is an important future work. Additionally, we require a modified Yang model to be used with NETCONF, which can support event notifications based on the radio measurement data, NGAP messages, etc. We expect that mobility management can be effectively handled by introducing a virtual OF switch (at MRN OF switch) for every UE. We aim to examine these open issues in the future.

\section{Acknowledgement}
This work has been supported by the Department of Telecommunications, Ministry of Communications, India, as part of the Indigenous 5G Test Bed project.



    \nocite{}	
	\bibliographystyle{IEEEtran}
	\bibliography{myrefs}

\begin{thebibliography}{10}
\providecommand{\url}[1]{#1}
\csname url@samestyle\endcsname
\providecommand{\newblock}{\relax}
\providecommand{\bibinfo}[2]{#2}
\providecommand{\BIBentrySTDinterwordspacing}{\spaceskip=0pt\relax}
\providecommand{\BIBentryALTinterwordstretchfactor}{4}
\providecommand{\BIBentryALTinterwordspacing}{\spaceskip=\fontdimen2\font plus
\BIBentryALTinterwordstretchfactor\fontdimen3\font minus
  \fontdimen4\font\relax}
\providecommand{\BIBforeignlanguage}[2]{{%
\expandafter\ifx\csname l@#1\endcsname\relax
\typeout{** WARNING: IEEEtran.bst: No hyphenation pattern has been}%
\typeout{** loaded for the language `#1'. Using the pattern for}%
\typeout{** the default language instead.}%
\else
\language=\csname l@#1\endcsname
\fi
#2}}
\providecommand{\BIBdecl}{\relax}
\BIBdecl

\bibitem{23.501}
\BIBentryALTinterwordspacing
``{3GPP TS 23.501, System Architecture for the 5G System},'' Tech. Rep., 2018,
  (Release 16). [Online]. Available:
  \url{https://www.3gpp.org/DynaReport/23501.htm}
\BIBentrySTDinterwordspacing

\bibitem{24.193}
\BIBentryALTinterwordspacing
``{3GPP TS 24.193, 5G System; Access Traffic Steering, Switching and Splitting
  (ATSSS)},'' Tech. Rep., 2020, (Release 16). [Online]. Available:
  \url{https://www.3gpp.org/DynaReport/24193.htm}
\BIBentrySTDinterwordspacing

\bibitem{openflowspec}
\BIBentryALTinterwordspacing
``{OpenFlow Switch Specification, v1.3.0},'' 2014. [Online]. Available:
  \url{https://www.opennetworking.org/wp-content/uploads/2014/10/openflow-spec-v1.3.0.pdf}
\BIBentrySTDinterwordspacing

\bibitem{softran}
A.~Gudipati, D.~Perry, L.~E. Li, and S.~Katti, ``{SoftRAN: Software defined
  radio access network},'' in \emph{Proceedings of the second ACM SIGCOMM
  workshop on Hot topics in software defined networking}, 2013, pp. 25--30.

\bibitem{softmobile}
T.~Chen, H.~Zhang, X.~Chen, and O.~Tirkkonen, ``{SoftMobile: Control evolution
  for future heterogeneous mobile networks},'' \emph{IEEE Wireless
  Communications}, vol.~21, no.~6, pp. 70--78, 2014.

\bibitem{flexran}
X.~Foukas, N.~Nikaein, M.~M. Kassem, M.~K. Marina, and K.~Kontovasilis,
  ``{FlexRAN: A flexible and programmable platform for software-defined radio
  access networks},'' in \emph{Proceedings of the 12th International on
  Conference on emerging Networking EXperiments and Technologies}, 2016, pp.
  427--441.

\bibitem{softnet}
H.~{Wang}, S.~{Chen}, H.~{Xu}, M.~{Ai}, and Y.~{Shi}, ``{SoftNet: A software
  defined decentralized mobile network architecture toward 5G},'' \emph{IEEE
  Network}, vol.~29, no.~2, pp. 16--22, 2015.

\bibitem{sdwn}
C.~J. Bernardos, A.~De~La~Oliva, P.~Serrano, A.~Banchs, L.~M. Contreras,
  H.~Jin, and J.~C. Z{\'u}{\~n}iga, ``{An architecture for software defined
  wireless networking},'' \emph{IEEE wireless communications}, vol.~21, no.~3,
  pp. 52--61, 2014.

\bibitem{softair}
I.~F. Akyildiz, P.~Wang, and S.-C. Lin, ``Softair: A software defined
  networking architecture for 5g wireless systems,'' \emph{Computer Networks},
  vol.~85, pp. 1--18, 2015.

\bibitem{softcell}
\BIBentryALTinterwordspacing
X.~Jin, L.~E. Li, L.~Vanbever, and J.~Rexford, ``{SoftCell: Scalable and
  Flexible Cellular Core Network Architecture},'' in \emph{Proceedings of the
  Ninth ACM Conference on Emerging Networking Experiments and Technologies},
  ser. CoNEXT ’13.\hskip 1em plus 0.5em minus 0.4em\relax New York, NY, USA:
  Association for Computing Machinery, 2013, p. 163–174. [Online]. Available:
  \url{https://doi.org/10.1145/2535372.2535377}
\BIBentrySTDinterwordspacing

\bibitem{openradio}
M.~Bansal, J.~Mehlman, S.~Katti, and P.~Levis, ``{Openradio: a programmable
  wireless dataplane},'' in \emph{Proceedings of the first workshop on Hot
  topics in software defined networks}, 2012, pp. 109--114.

\bibitem{orchestra}
T.~{De Schepper}, P.~{Bosch}, E.~{Zeljković}, F.~{Mahfoudhi},
  J.~{Haxhibeqiri}, J.~{Hoebeke}, J.~{Famaey}, and S.~{Latré}, ``Orchestra:
  Enabling inter-technology network management in heterogeneous wireless
  networks,'' \emph{IEEE Transactions on Network and Service Management},
  vol.~15, no.~4, pp. 1733--1746, Dec 2018.

\bibitem{empower}
E.~{Coronado}, S.~N. {Khan}, and R.~{Riggio}, ``{5G-EmPOWER: A Software-Defined
  Networking Platform for 5G Radio Access Networks},'' \emph{IEEE Transactions
  on Network and Service Management}, vol.~16, no.~2, pp. 715--728, June 2019.

\bibitem{openroads}
K.-K. Yap, M.~Kobayashi, R.~Sherwood, T.-Y. Huang, M.~Chan, N.~Handigol, and
  N.~McKeown, ``{OpenRoads: Empowering Research in Mobile Networks},''
  \emph{SIGCOMM Comput. Commun. Rev.}, vol.~40, no.~1, p. 125–126, Jan. 2010.

\bibitem{openroads2}
K.-K. Yap, M.~Kobayashi, D.~Underhill, S.~Seetharaman, P.~Kazemian, and
  N.~McKeown, ``{The stanford openroads deployment},'' in \emph{Proceedings of
  the 4th ACM international workshop on Experimental evaluation and
  characterization}, 2009, pp. 59--66.

\bibitem{openran}
M.~Yang, Y.~Li, D.~Jin, L.~Su, S.~Ma, and L.~Zeng, ``{OpenRAN: a
  software-defined ran architecture via virtualization},'' \emph{ACM SIGCOMM
  computer communication review}, vol.~43, no.~4, pp. 549--550, 2013.

\bibitem{sdn_akshatha}
A.~N. {Manjeshwar}, A.~{Roy}, P.~{Jha}, and A.~{Karandikar}, ``{Control and
  Management of Multiple RATs in Wireless Networks: An SDN Approach},'' in
  \emph{IEEE 2nd 5G World Forum (5GWF)}, 2019, pp. 596--601.

\bibitem{open5g}
P.~K. Taksande, P.~Jha, A.~Karandikar, and P.~Chaporkar, ``{Open5G: A
  Software-Defined Networking Protocol for 5G Multi-RAT Wireless Networks},''
  in \emph{Wireless Communication and Networking Conference}, 2020.

\bibitem{multiratmc}
S.~{Chandrashekar}, A.~{Maeder}, C.~{Sartori}, T.~{Höhne}, B.~{Vejlgaard}, and
  D.~{Chandramouli}, ``{5G multi-RAT multi-connectivity architecture},'' in
  \emph{2016 IEEE International Conference on Communications Workshops (ICC)},
  2016, pp. 180--186.

\bibitem{lwa1}
D.~{Laselva}, D.~{Lopez-Perez}, M.~{Rinne}, and T.~{Henttonen}, ``{3GPP
  LTE-WLAN Aggregation Technologies: Functionalities and Performance
  Comparison},'' \emph{IEEE Communications Magazine}, vol.~56, no.~3, pp.
  195--203, 2018.

\bibitem{lwa2}
P.~{Nuggehalli}, ``{LTE-WLAN aggregation [Industry Perspectives]},'' \emph{IEEE
  Wireless Communications}, vol.~23, no.~4, pp. 4--6, 2016.

\bibitem{netconf}
R.~Enns, M.~Bjorklund, J.~Schoenwaelder, and A.~Bierman, ``{Network
  configuration protocol (NETCONF), RFC 6241},'' 2011.

\bibitem{openflowconfig}
\BIBentryALTinterwordspacing
``{OpenFlow Management and Configuration Protocol, v1.2},'' 2014. [Online].
  Available:
  \url{https://www.opennetworking.org/images/stories/downloads/sdn-resources/onf-specifications/openflow-config/of-config-1.2.pdf}
\BIBentrySTDinterwordspacing

\bibitem{openflownotifications}
\BIBentryALTinterwordspacing
``{OpenFlow Notifications Framework, OpenFlow Management, v1.0},'' 2013.
  [Online]. Available:
  \url{https://www.opennetworking.org/wp-content/uploads/2013/02/of-notifications-framework-1.0.pdf}
\BIBentrySTDinterwordspacing

\bibitem{matlab}
\BIBentryALTinterwordspacing
``{MATLAB 2020a}.'' [Online]. Available: \url{www.mathworks.com}
\BIBentrySTDinterwordspacing

\bibitem{code}
\BIBentryALTinterwordspacing
``{5G-Flow Simulator Source Code}.'' [Online]. Available:
  \url{https://github.com/MeghnaKhaturia/5G-Flow-RAN}
\BIBentrySTDinterwordspacing

\bibitem{38.901}
\BIBentryALTinterwordspacing
{3GPP TR 38.901}, ``{Study on channel model for frequencies from 0.5 to 100 GHz
  (Release 15)},'' Tech. Rep., 2018. [Online]. Available:
  \url{https://www.3gpp.org/DynaReport/38.901.htm}
\BIBentrySTDinterwordspacing

\bibitem{802.11model}
\BIBentryALTinterwordspacing
{IEEE 802.11-14/0882r4}, ``{IEEE 802.11ax Channel Model Document},'' Tech.
  Rep., 2014. [Online]. Available:
  \url{https://mentor.ieee.org/802.11/dcn/14/11-14-0882-04-00ax-tgax-channel-model-document.docx}
\BIBentrySTDinterwordspacing

\bibitem{802.11}
``{IEEE Standard for Information technology—Telecommunications and
  information exchange between systems Local and metropolitan area
  networks—Specific requirements - Part 11: Wireless LAN Medium Access
  Control (MAC) and Physical Layer (PHY) Specifications},'' \emph{{IEEE Std
  802.11-2016 (Revision of IEEE Std 802.11-2012)}}, pp. 1--3534, Dec 2016.

\bibitem{38.331}
\BIBentryALTinterwordspacing
``{3GPP TS 38.331, 5G; NR; Radio Resource Control (RRC); Protocol
  specification},'' Tech. Rep., 2020, (Release 16). [Online]. Available:
  \url{https://www.3gpp.org/dynareport/38331.htm}
\BIBentrySTDinterwordspacing

\end{thebibliography}

\end{document}